\documentclass[preprint,12pt]{elsarticle}

\usepackage{amssymb}
\usepackage{amsmath}
\usepackage{graphicx}
\usepackage{dcolumn}
\usepackage{bm}

\usepackage[utf8]{inputenc}
\usepackage[T1]{fontenc}
\usepackage{mathptmx}
\usepackage{etoolbox}
\usepackage{subcaption}
\usepackage{ar}
\usepackage{xcolor}
\usepackage{amsthm}
\usepackage[justification=centering]{caption}
\usepackage{array}
\usepackage{booktabs}
\usepackage{lscape}

\journal{Flow Measurement and Instrumentation}

\begin{document}

\begin{frontmatter}



\title{Performance Evaluation of a Fan-Array Wind Tunnel for Advanced Aerodynamic Studies}


\author{Kalyani Panigrahi}
\author{Rohan Bhattacharya}
\author{Sabareesh G.R.}
\author{Pardha S Gurugubelli}
\affiliation{Department of Mechanical Engineering,
BITS Pilani Hyderabad Campus, Hyderabad, 500078, Telangana, India}
\begin{abstract}
Replicating and analyzing complex real-time flow conditions presents significant challenges for aerodynamicists. Wind tunnel's consisting an array of computer controlled/programmable fans have evolved over the last three decades for simulating the atmospheric turbulence to study the aerodynamic loads over bluff bodies such as buildings and bridges. In addition to civil engineering structures, these wind tunnels can also have applications extending towards free-flight aerodynamic tests on drones and unmanned aerial vehicles (UAVs). Achieving velocity profiles such as linearly sheared flows or boundary layer flows in a traditional wind tunnel would require long fetch lengths for the flow to evolve. However, in this work, with the aid of an array of computer-controlled fans, a perfectly linearly sheared flow can be developed at a distance of 1 meter from the source. Unlike the previous fan-array wind tunnels (FAWT) developed for atmospheric boundary layer flows, the fan-array tunnel constructed in this work does not include a contraction section or a honeycomb, offering portability and compactness for performing unconfined flight tests.
The current work involves the design and development of a 10 × 10 two-dimensional array of multiple fans facing a 2.4 m long, enclosed test section with a 1.44 $m^2$ cross-sectional area. A comprehensive characterization of key flow parameters, including turbulence intensity, mean velocity, and pressure, in the streamwise, spanwise, and transverse directions of the test section is investigated in the developed FAWT. These measurements were conducted under three distinct flow conditions, namely uniform, linearly sheared, and parabolic. Additionally, this work also presents the evolution of the turbulent length scales and time scales across various sections inside the tunnel for a range of wind speeds by employing autocorrelation techniques and power spectral analysis.
\end{abstract}
\end{frontmatter}



\section{Introduction}
Advanced aerodynamic research is a dynamic and multifaceted field, and is driven by the demand for more efficient and sustainable aerial and ground-based systems. The studies include actively manipulating the flow around a structure, leading to improved performance in designing aircraft for urban environments such as Urban Air Mobility (UAM), Electric Vertical Take-off and Landing (eVTOL), and Micro/Nano-Air vehicles (MAVs/NAVs). Light, robust, yet aerodynamically efficient vehicles are essential for applications such as surveillance, search and rescue, military and defense, and communication networks, among other uses. The design and construction of these devices incorporate aerodynamic optimization, drag reduction, and autonomous control systems. These systems are developed to efficiently handle complex real-life wind conditions (Cook \cite{cook1982}, Holmes \cite{holmes2007wind}, Wang et al. \cite{wang1a2023systematic}, Lee et al. \cite{lee2023updates}, Adhikari et al. \cite{adhikari2023extreme}), thereby necessitating either wind tunnel tests or numerical experiments that accurately replicate atmospheric conditions. However, most of the numerical works do not take the effect of unsteady wind loads on the stability of an aerial vehicle (Park et al. \cite{park2023numerical}) primarily due to the computational challenges associated with generating unsteady wind conditions (Daemei et al. \cite{Daemei2018}; Shen et al. \cite{shen2022non}; Hao et al. \cite{hao2017nonsynoptic,hao2018downburst}; Jing et al. \cite{jing2023study}; Ibrahim et al. \cite{Ibrahim2020}). The need to verify and test these devices under unsteady and turbulent flow conditions (Kopp \cite{kopp2023updates}) has spurred the evolution of wind tunnel technologies. This has led to a shift from traditional methods using grids and surface blocks with long test sections (Wani et al. \cite{wani2021experimental}; Chen et al. \cite{chen2023unsteady}) to compact, computer-controlled arrays of small fans that can precisely generate desired velocity profiles (Li et al. \cite{li2024aerodynamic}). 

One of the early implementations of using multiple fans consisted of 11 rows and 6 columns of AC induction motor-driven fans to induce air flow can be found in the works presented by Nishi et al. \cite{nishi1993computer} at Miyazaki University of Japan. The work focuses on generating boundary layer velocity profiles inside a 5.7 m long and 1 m x 1 m test-section. The fan-array driven wind tunnel facility includes a contraction section and a honeycomb core placed at the outlet of the contraction section, with each cell 10 mm in diameter and 100 mm long. The authors have measured the turbulence intensities (TI) and the mean streamwise velocity inside the tunnel at a section 3 m from the inlet to the test-section and have shown that the TI value is about 2-7 \% for a range of mean wind speeds ranging between 5-10 m/s and velocity profiles. Nishi et al. continued their work on multiple fan wind tunnels to develop a closed-loop feedback system for controlling the fan-array system to generate the desired flow characteristics such as TI, length scales, mean velocity profile, etc. (Nishi et al. \cite{nishi1995computer,nishi1997turbulence}, Cao et al. \cite{shuyang2001actively,cao2002reproduction}). 

Another study at the Insurance Institute of Business and Home Safety (IBHS) research facility in Richburg (Smith et al. \cite{smith2012simplified}) developed a fan-array tunnel with a closed loop PID based control system for controlling the 3 × 7 array of propeller fans that push air into a 4.9 m long, 4.6 m wide and 1.8 m high test section through a 2.9 m wide and 0.9 m high contraction section. They have demonstrated that the fan array can generate boundary layer profiles corresponding to the marine, open, and suburban conditions. S. Ozono et al. \cite{ozono2006turbulence} constructed a fan-array facility with a contraction section powered by an 11 × 9 array of AC Servo motors. The tunnel is provided with a honeycomb at the exit of the contraction section, and the dimensions of the test section are 2.54 m wide and 1.80 m high. In this work, the authors have investigated the impact of test-section size and fan-array operation grid pattern on the turbulence characteristics and mean velocity profiles. They extended their work to generate isotropic high Reynolds number turbulence \cite{ozono2007turbulence} by using the combination of the previously applied modes. The downstream development of the root mean squared (rms) velocity, anisotropy, mean velocity profile, turbulence characteristics, and the power spectrum on grid operation pattern were compared and investigated. In \cite{ozono2018turbulence}, the authors have managed to control the integral length scales of the flow by controlling the fan RPMs when operated in a staggered grid. 


Most of the fan arrays discussed above have been developed to generate atmospheric boundary layer flows, however Johnson and Jacob \cite{johnson2009development} constructed a fan-array tunnel consisting of a 11 × 10 fan-array having a test cross-section of 1 × 1 $m^2$ to investigate the behaviour of MAVs and NAVs when exposed to sudden gusts and shear flows using velocity contour plots. The tunnel is provided with a honeycomb structure placed downstream of the fans. The maximum velocity in the tunnel was measured to be 5 m/s without the flow straightener and a slightly lesser value with it present. Velocity profiles along the width and height of the tunnel were presented by performing particle image velocimetry (PIV) measurements. The velocity fluctuations were measured to be in the order of 20 cm/s or $\approx$ 5\% of the mean flow. Dougherty \cite{dougherty2018fan} from Caltech University constructed a fan-array tunnel facility consisting of 44 x 44 fans for attaining a maximum velocity of 7.5 m/s. The tunnel was constructed to study the flow patterns of uniform, shear, vortex, and gusting flows. They have presented the TI and free-stream mean velocity measurements in the longitudinal directions, with and without the presence of honeycomb structures downstream. The TI values range from $\approx$ 8\% to 60\% without the honeycombs and $\approx$ 3\% to 55\% with the honeycombs for uniform flow mean velocities ranging between 1 to 12 m/s. The effect of fan structure or the measurement location is no longer observed beyond the non-dimensional distance $x/L=$ 0.5. 


One of the most recent studies includes the aerodynamic characterization of a 40 × 40 $cm^2$ unbounded fan-array containing 100 fans \cite{li2024aerodynamic} using hot-wire anemometer (HWA) and PIV measurements in the streamwise directions for calculating turbulence intensities (TI) and mean stream velocities for four distinct planes at distances ranging from 4 to 16 cm from the fan-arrays and for duty-cycles $\in$ 50\%, 80\%, 100\%. They observed a TI value of less than 10\% for approximately 36\% of the total test cross-sectional area when operated at 100\% duty cycle, and this area of less than 10\% drops below 25\% for a duty cycle of 50\%. Another work by Liu et al. \cite{Liu2025} presents the aerodynamic characterization of a 40 x 40 fan array wind generator (FAWG) having a 3.25 x 3.25 $m^2$ test-section with the freestream velocity of 13 m/s, by employing an 8 x 8 equidistant pitot tube array and HWA for detailed flow characterization. The aerodynamic characteristics of another multi-fan configuration employed to produce airflow in a 0.8 x 1.5 $m^2$ test section at a freestream velocity of 6 m/s have been investigated by Chaudhary et al. \cite{Chaudhary2025}. Another work by Di Luca et al. \cite{DiLuca2024a} presented an integration of a flow management device into a fan-array wind tunnel along with 3 honeycombs and 4 stainless steel mesh screens to achieve lower turbulence levels, mostly observed in traditional single impeller wind tunnels (0.45\% – 7\%) while reducing the momentum output. He later derived an analytical model (\cite{DiLuca2024b}) for developing shear flows with flow management devices employed in the FAWT, which has been validated through PIV measurements of a small-scale FAWT. 

In a study, Veismann et al. \cite{veismann2021low} recreated the Martian conditions by using an open-jet multi-fan wind tunnel comprising 441 individually controlled fan units. This arrangement was integrated into the Jet Propulsion Laboratory (JPL) facility of NASA to carry out free and tethered forward flight testing in a Martian environment. They later extended their work by investigating the performance of a variable pitch multi-rotor in an external upflow of axial descent at the Centre of Autonomous Systems and Technologies (CAST) at Caltech \cite{Veismann2021axial}. This facility is composed of 1296 DC fan units having a test-section of 2.88 x 2.88 $m^2$ cross-sectional area with the wind velocities varying from 0-14 m/s. The fan units were operated under identical conditions, thereby creating uniform flow conditions, and could be oriented both horizontally and vertically. A recent work by Walpen et al. \cite{walpen2023real} designed a weather simulator having a large number of fans that can be arranged in various configurations to recreate desired wind and weather conditions. They have performed the flow characterization of the wind shaper by using a 3-axis hot-wire anemometer and have also reported atmospheric turbulence at drone scales. This work was later extended to fabricating an automated control and feedback mechanism for the collection and processing of real-time flow data, thereby reducing the time and offering flexibility while controlling complex aerodynamic flow conditions \cite{walpen2024automated}. Zavala et al. \cite{stefan2024data} investigated constant fan speeds and time-averaged streamwise velocities by building a simple predictive tool operating under a limited set of conditions and with a basic control strategy.


Every wind tunnel is unique and distinctive due to its design, size, and flow parameters (Eckhert \cite{eckert1976aerodynamic}, Mehta and Bradshaw \cite{mehta1979design}). From the above literature, it can be concluded that fan-array wind tunnels (FAWT) provide control and flexibility to tune the flow parameters inside the tunnel, creating diverse wind conditions. Consequently, there is a growing demand for advanced wind tunnels worldwide. These facilities must be capable of generating complex wind profiles and gusts across a wide range of turbulence intensities and scales for the comprehensive aerodynamic testing of unmanned aircraft. From the above works, it is clear that FAWTs have the capability to tune the flow field both spatially and temporally by manipulating the fan RPMs using a computer program with a closed-loop feedback system. These capabilities highlight their importance, particularly given the critical need for robust and versatile facilities to generate complex wind and gust profiles for the development of modern flight control systems. Understanding aerodynamic characteristics of FAWTs for different operating conditions is crucial for the development/training of machine learning algorithms that can develop desired wind conditions.


The current work presents the design, development, and aerodynamic characterization of an FAWT consisting of a fan array with 10 rows and 10 columns. The FAWT developed consists of a 2.4 m long test-section that is enclosed on four sides and has a cross-sectional area of 1.2 × 1.2 $m^2$. Similar to the FAWTs developed for testing of UAVs/MAVs (\cite{johnson2009development,dougherty2018fan, Veismann2021axial,li2024aerodynamic}, the FAWT developed as part of this work also does not contain either a contraction section or a honeycomb. Each of these fans can be independently controlled and programmed using a microcontroller. In this work, the aerodynamic characteristics of the tunnel are determined by evaluating the turbulence intensity (TI), mean velocity, turbulent length, and time scales by employing a hot-wire anemometer (HWA) mounted on an automated traverse system controlled by a microcontroller. Additionally, the work also reports the variation of the static and total pressure inside the test-section, measured using a pressure rake that is connected to a 16-port pressure scanner.
The content of this paper is organized as follows: Section 2 presents the experimental setup and the procedure used to determine the aerodynamic characteristics of the FAWT developed, and Sections 3 and 4 report the different aerodynamic characteristics observed for uniform and non-uniform flow conditions.

\section{Experimental Setup and Measurement Techniques}
\subsection{Fan-array wind tunnel design}
\begin{figure}
\centering
\includegraphics[width=\textwidth,trim={0cm 0cm 0cm 0},clip]{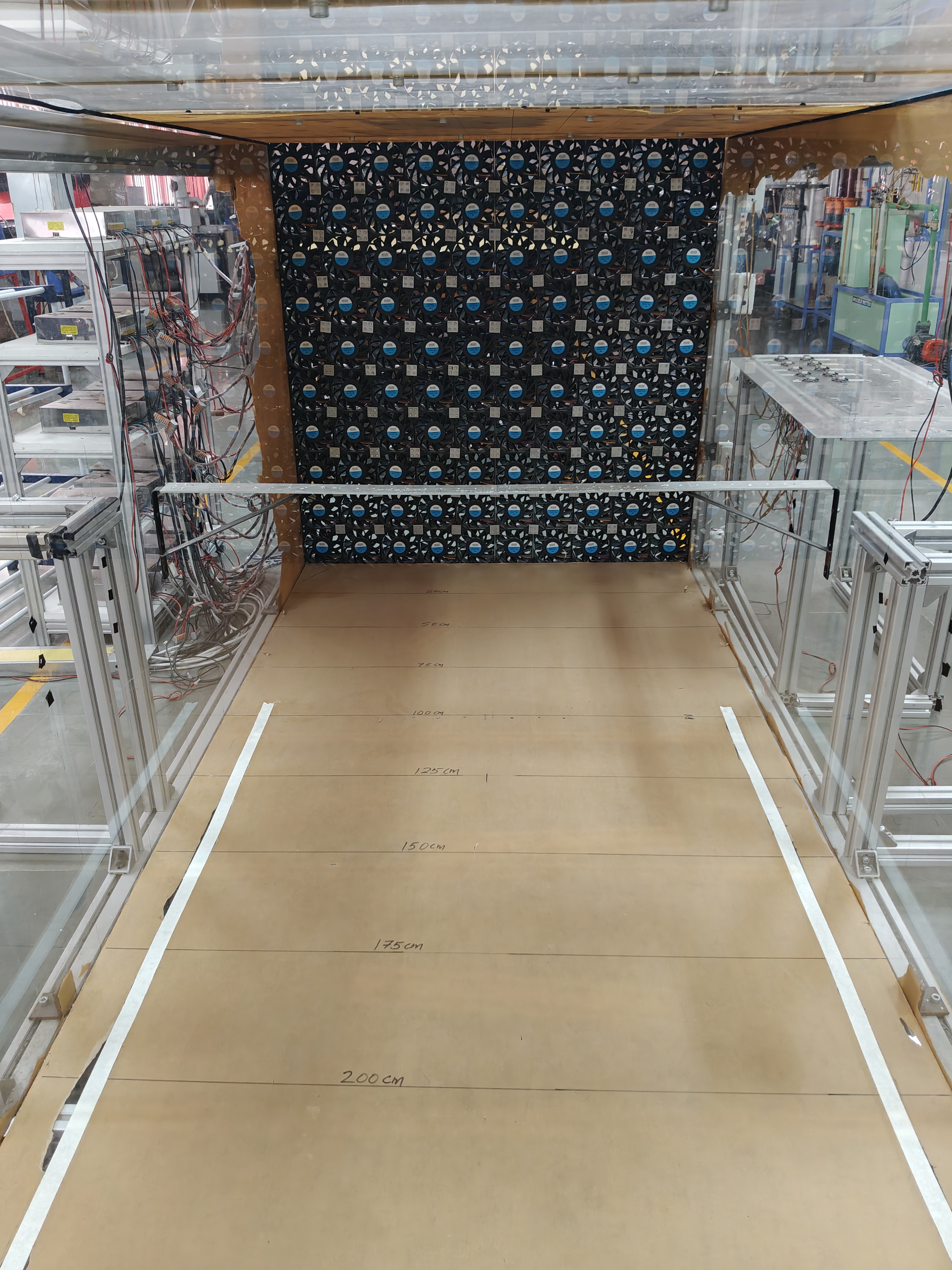}
\caption{The 10 x 10 Fan-Array Wind Tunnel Facility}
\label{fig:FAWTpic}
\end{figure}

In this work, a 2.4 m long fan-array wind tunnel (FAWT) with a cross-sectional area of 1.2 × 1.2 $m^2$ is constructed as shown in Figure \ref{fig:FAWTpic}. The test section is open-ended and enclosed with transparent acrylic walls on four sides to ensure clear visibility. The tunnel is powered by 10 rows and 10 columns of BLDC (Brushless DC) fans arranged in a two-dimensional array-like structure. These are Hicool four-wire DC compact fans that operate at a 48V DC power supply, requiring an input current of 1330 mA. The fans are arranged in a square grid pattern, and each fan covers an area of 120 × 120 $mm^2$, as shown in Figures \ref{fig:FAWTpic} and \ref{fig:Schematic}. The maximum air speed achievable from each fan is 5500 rpm with a maximum flow rate of 285 cubic feet/min (CFM), and the RPM of the fan can be controlled by providing a 5V pulsed signal at 25 kHz. The RPM of each fan has been tested, and it has been ensured that the RPM is directly proportional to the width of the pulsed signal generated. 

Each port is 50 mm long with an outer diameter of 2.18 mm. A 16-port pressure rake was developed to thoroughly examine pressure characteristics across the test section. The rake is mounted on a traverse system and will be used to measure static and total pressure at various points within the test section. The multiple ports are positioned at equidistant heights along the The pressure rake is connected to a pneumatic pressure scanner (Netscanner Model 9216), which samples up to 500 measurements per channel per second, as shown in Figure \ref{fig:rake}. To ensure repeatability and rule out any experimental uncertainty, each run is repeated 10 times at each location.

\subsection{Flow Measurement and Control Techniques}
\begin{figure}
\centering
\begin{subfigure}[b]{0.5\textwidth}
\centering
\includegraphics[width=\textwidth,trim={4.5cm 0.6cm 9cm 0.4cm},clip]{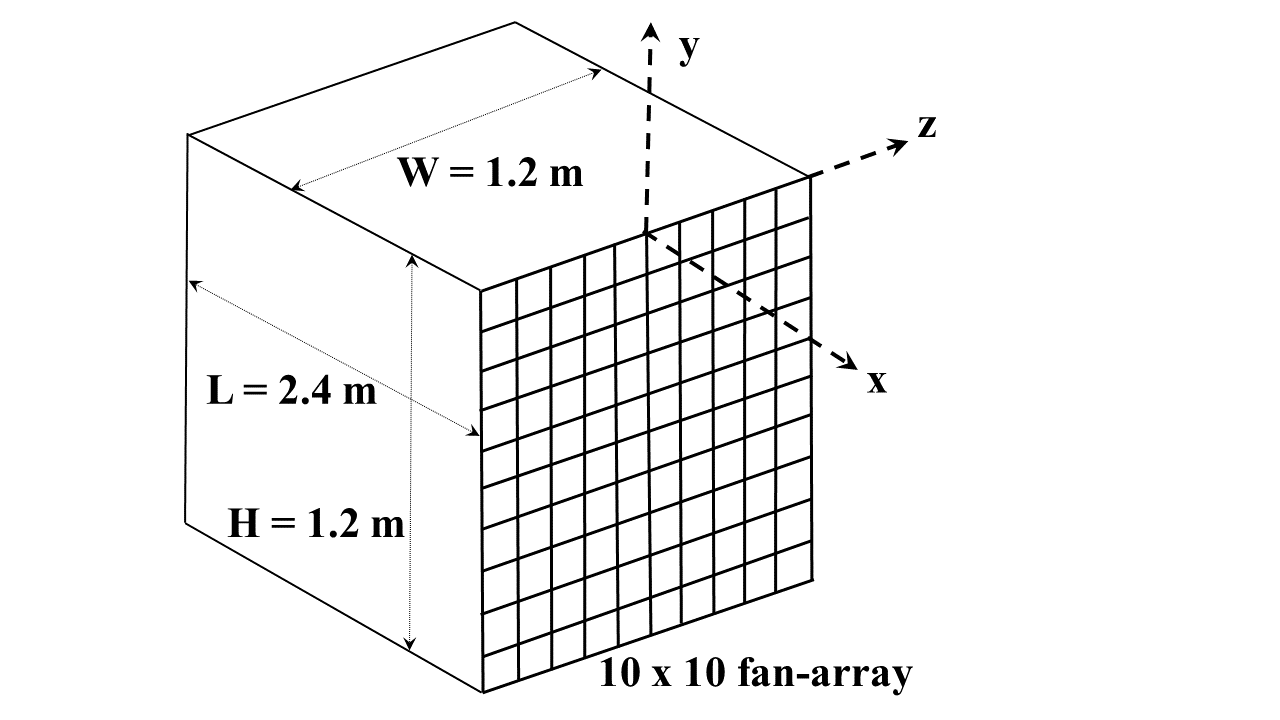}
\caption{ }
\label{fig:Schematic}
\end{subfigure}
\hfill
\begin{subfigure}[b]{0.49\textwidth}
\centering
\includegraphics[width=\textwidth,trim={0.2cm 0cm 0cm .2cm0},clip]{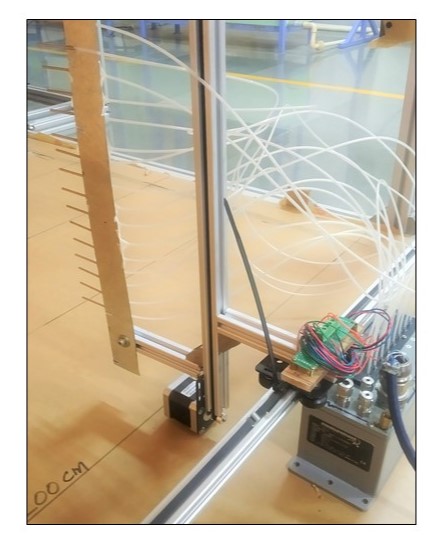}
\caption{ }
\label{fig:rake}
\end{subfigure}
\caption{ (a) A schematic representation of the fan-array with the x-y-z configuration (b) Multi-port pressure rake used for pressure measurements} \label{schematicandrake}
\end{figure}

Figure \ref{fig:Schematic} shows a 3-dimensional schematic representation of the FAWT, with the flow characteristics measured along the $x,y,z$ directions along the Length ($L$), Height ($H$), and Width ($W$) of the tunnel, respectively. To investigate detailed flow characterization inside the tunnel, a Dantec Dynamic fiber constant temperature hot-wire anemometer (HWA) mounted on an automated traverse system is employed to capture the time histories of the mean free stream velocity ($\overline{U_0}$) and turbulence fluctuations (TI) at various locations inside the test-section along the longitudinal and transverse directions as shown in figure \ref{fig:Schematic}. This approach allows the data acquisition to be accomplished with high temporal resolution. The HWA consists of a fibre probe with straight prongs, with the sensor perpendicular to the probe axis (Make$:-$Dantec Dynamics, Model$:-$ 55R01). To minimize the blockage effects the HWA is mounted onto a 22.5 cm long probe holder made of stainless steel with a 4 mm diameter (Make$:-$Dantec Dynamics, Model$:-$ 55H21), which is affixed to a traverser operated by a simple belt and pulley arrangement for precise and smooth positioning of the fiber probe at desired locations across the working section of the FAWT. The calibration of the hot-wire was performed by establishing a correlation between the voltages acquired by the HWA and the corresponding known freestream velocities (measured by a Pitot Tube earlier) by using a 5th-order polynomial for curve fitting to model the relationship between the voltages and the velocities. The HWA is connected to a 4-channel analog input module (NI-9215) with a maximum sampling rate of 100 ks/s/channel for data acquisition. NI devices are an obvious choice for data collection, due to their accuracy, performance, and control flexibility. It is operated at a 10 kHz sampling rate with a 10-second data acquisition duration. MATLAB software is utilized for data analysis and representation.

\begin{figure}
\centering
\includegraphics[width=\textwidth,trim={1.5cm 0cm 3.5cm 0.5cm},clip]{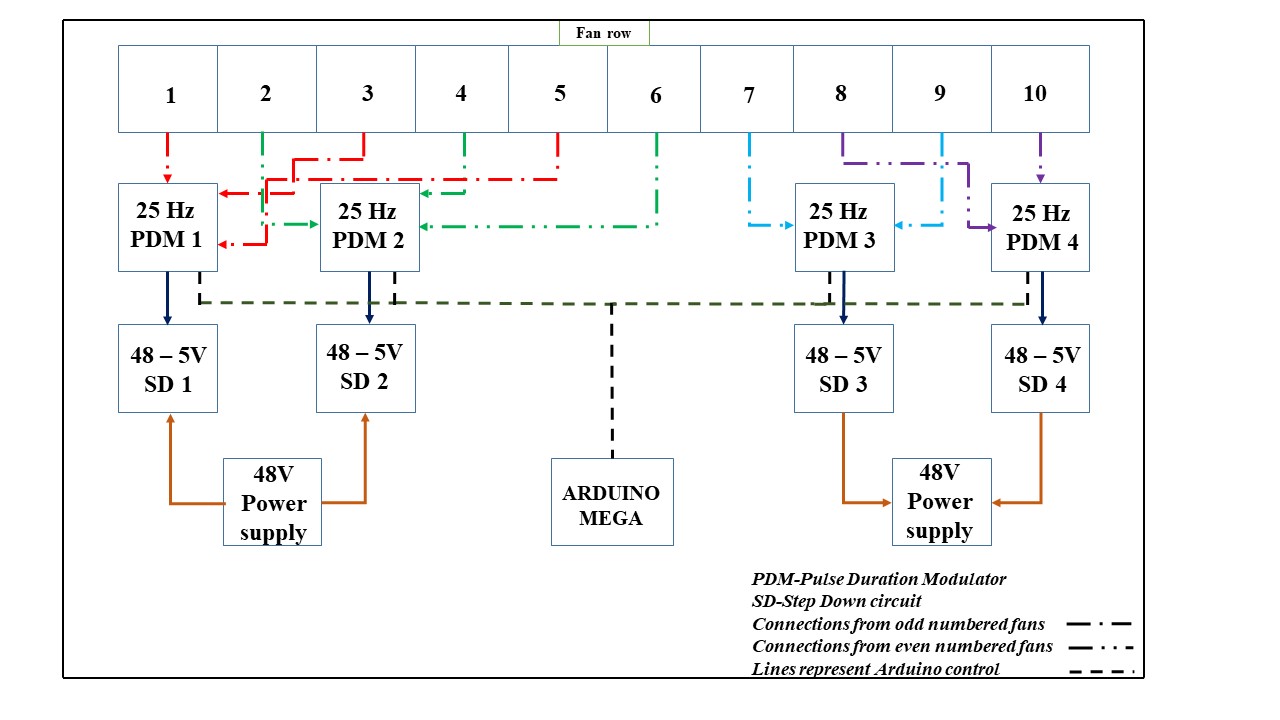}
\caption{Schematic layout depicting the electrical control of a single row of fans}
\label{fig:layout}
\end{figure}

The schematic of the electrical power and control wiring layout of the FAWT is presented in Figure \ref{fig:layout}. The 48V DC power supply to the fans in a given row is provided by two switched-mode power supplies (SMPS) rated at 10A. Each SMPS is connected to 5 alternating fans as shown in figure \ref{fig:layout}. The speed (RPM) of the fans is controlled using a programmable pulse duration modulator (PDM), and each PDM can control a maximum of 5 fans owing to its current rating. The output voltage of the SMPS is rated at 48 V, whereas the required voltage for the PDM operation is 5V. Therefore, the 5V input for the PDM is generated by using a 48V to 5V DC-DC step-down converter having a rated current of 2A. A total of 20, 2-channel square wave generator PDM modules are employed to individually control the DC fans in the 10 x 10 array. The fans are operated at a constant frequency of 25 kHz, while their duty cycles are adjusted from 1-100\%. The fan speed varies linearly with the duty cycles. Crucial to the fan array's operation, power regulation has been implemented to provide flexible control and the ability to detect faults in individual fans.
The microcontroller considered for the current work is an Arduino Mega 2560 board, and it can set the desired pulse width/duration that needs to be generated by a PDM by sending a serial command. With the setup configuration, the measurement and control techniques elaborated in this section, the FAWT is capable of generating controlled wind environments with both temporal and spatial precision. The variation of the flow parameters for the uniform and non-uniform flow conditions is extensively discussed in the upcoming sections.


\section{Uniform Velocity Profiles}
To achieve a wide spectrum of real-time flow conditions, the fans are operated at uniform and non-uniform operating modes. In the uniform mode, all the 100 fans are operated at a single duty cycle, with the range of duty cycles varying from 10 to 100\%. At these operating conditions, the flow parameters are investigated in detail. 
This section presents the variation of the free stream velocity ($\overline{U_0}$) and turbulent characteristics, such as TI, turbulent kinetic energy spectrum, turbulent length and time scales ($l_s$ and $\bar{\tau}$) along the non-dimensional length ($x/L$), the non-dimensional height ($y/H$), and the non-dimensional width ($z/W$) for a range of operating speeds. Here, $x$ represents the longitudinal distance from the fan array, $y$ is the height along the test section, and $z$ represents the transverse locations along the width of the tunnel. $L$, $H$, and $W$ represent the length, height, and width of the test-section, respectively, as shown in figure \ref{fig:Schematic}. Additionally, this section would also discuss the total and static pressure distribution ($P_{Total}$, $P_{Static}$) across various sections of the wind tunnel obtained from the pressure scanner.

\subsection{Mean velocity and Turbulence Fluctuations}

\begin{figure}
\centering
\begin{subfigure}[b]{0.49\textwidth}
\centering
\includegraphics[width=\textwidth,trim={4.5cm 0.6cm 9cm 0.4cm},clip]{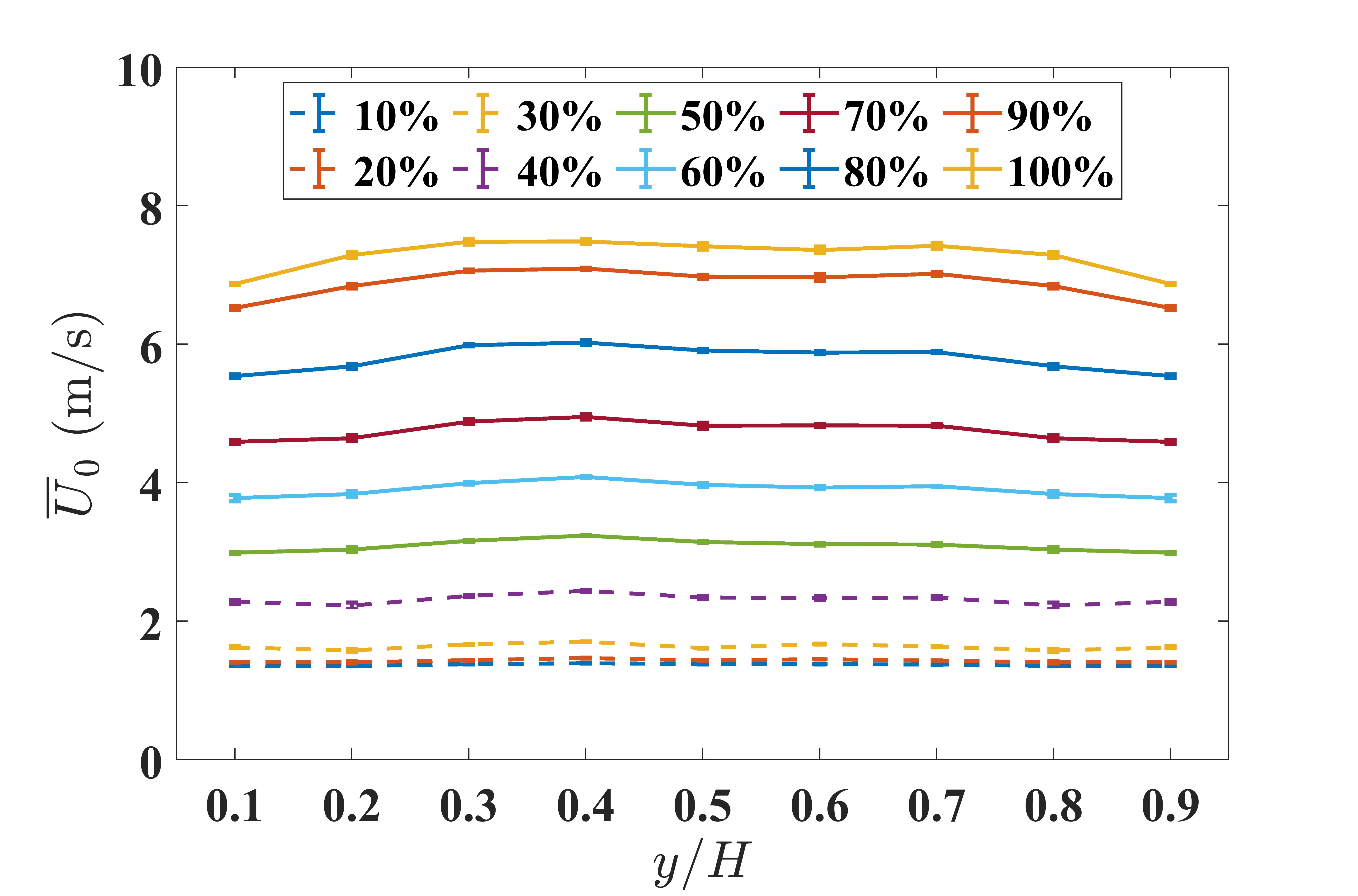}
\caption{ }
\label{fig:umeanvsy}
\end{subfigure}
\hfill
\begin{subfigure}[b]{0.49\textwidth}
\centering
\includegraphics[width=\textwidth,trim={4.5cm 0.6cm 9cm 0.4cm},clip]{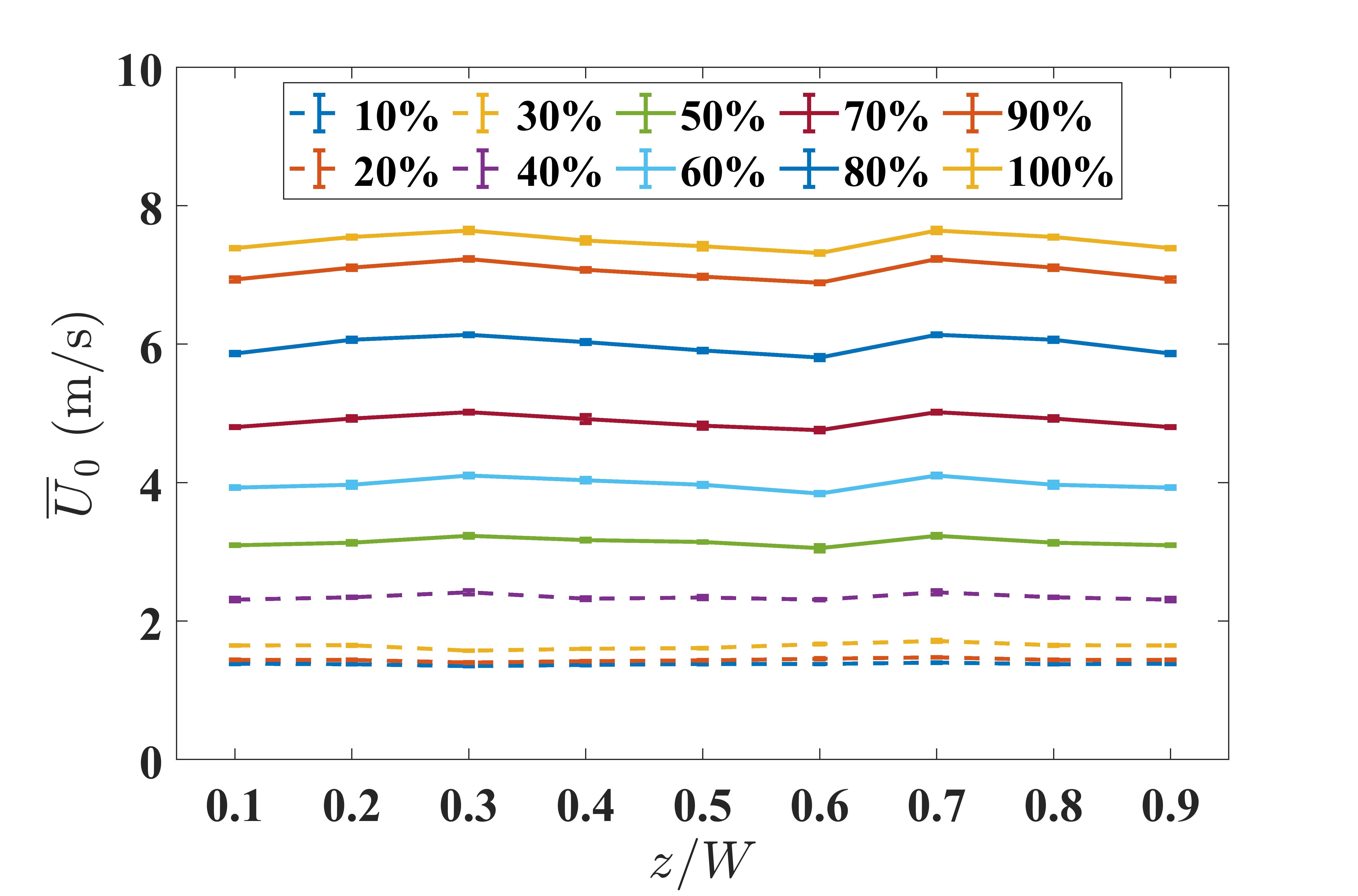}
\caption{ }
\label{fig:umeanvsz}
\end{subfigure}
\caption{ Variation of mean velocity ($\overline{U_0}$) along the (a) non-dimensional height ($y/H$), and (b) non-dimensional width ($z/W$) as a function of PDM duty-cycle for the cross-section located at $x/L = 0.52$.}
\label{Umeanvsyandz}
\end{figure}

To analyze $\overline{U_0}$ across different locations inside the test-section, the streamwise velocity component is measured for a cross-section located at a distance $x/L$ = 0.52 from the fan-array using the HWA at nine points along the height $y/H = \{0.1,0.2, 0.3, 0.4, 0.5, 0.6, 0.7, 0.8, 0.9\}$ at  $z/W$ = 0.5 and nine points 
$z/W = [0.1 $-$ 0.9]$ along the width at $y/H$ = 0.5. Figures \ref{fig:umeanvsy} and \ref{fig:umeanvsz} present the variation of $\overline{U_0}$ along the height and width, respectively, for a range of BLDC fans’ duty-cycles starting from 10\% to 100\%. It can be observed that the $\overline{U_0}$ remains mostly uniform across the cross-section for a given duty cycle. The $\overline{U_0}$ values at the center of the test-section increase linearly for duty-cycles between 30\% and 90\%. It remains constant at a cross-section for $y/H$, $z/W$ $\in [0.2 $-$ 0.8]$. The boundary-layer effect has been observed for $y/H$, $z/W\le$ 0.2 and $y/H$, $z/W\ge$ 0.8. The error bar in Figures \ref{fig:umeanvsy} and \ref{fig:umeanvsz} represents the standard deviation of $\overline{U_0}$, and the maximum standard deviation that has been observed is less than 0.03 m/s.

\begin{figure}
\centering
\begin{subfigure}[b]{0.49\textwidth}
\centering
\includegraphics[width=\textwidth,trim={4.5cm 0.6cm 9cm 0.4cm},clip]{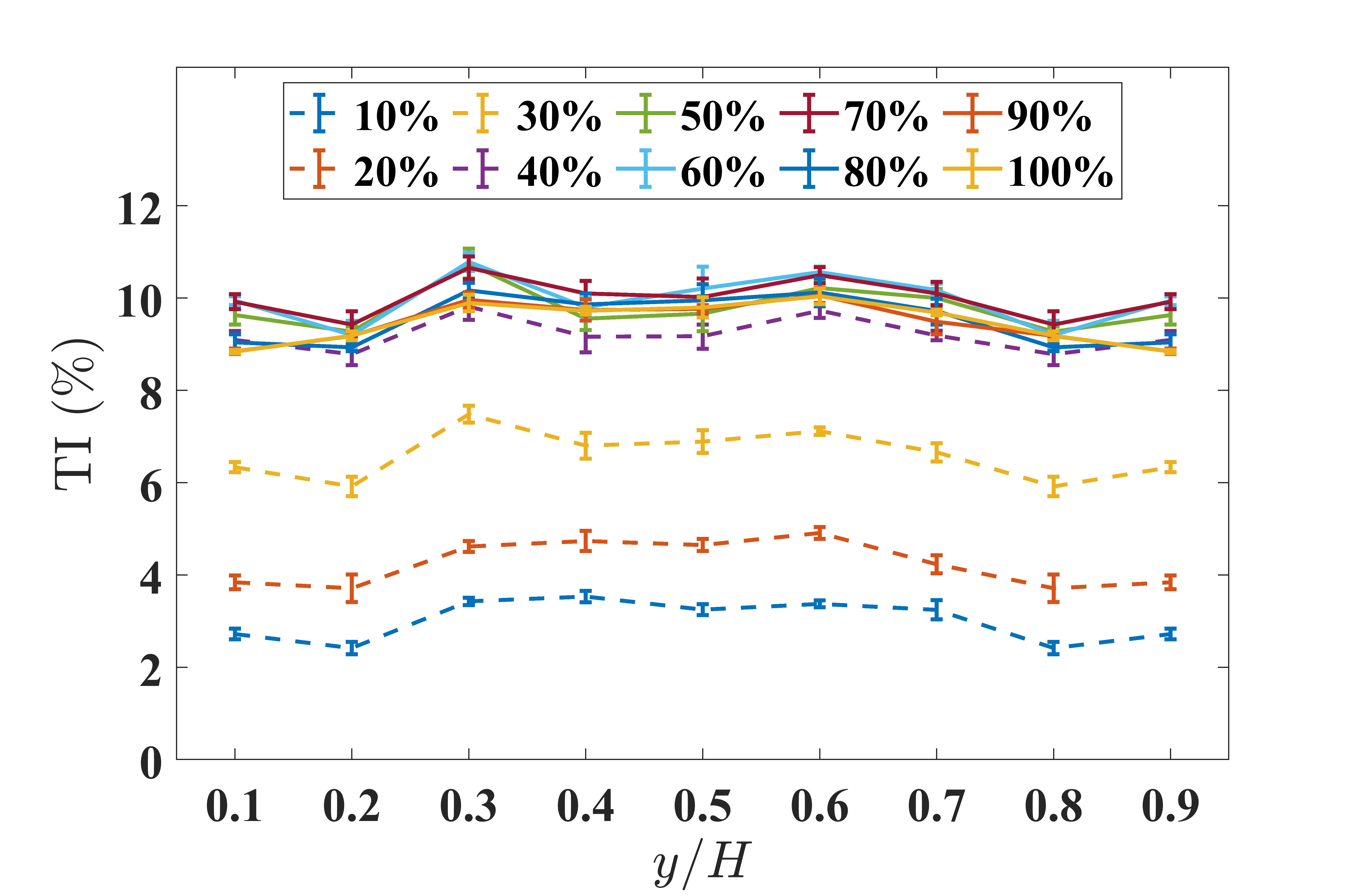}
\caption{ }
\label{fig:tivsy}
\end{subfigure}
\hfill
\begin{subfigure}[b]{0.49\textwidth}
\centering
\includegraphics[width=\textwidth,trim={4.5cm 0.6cm 9cm 0.4cm},clip]{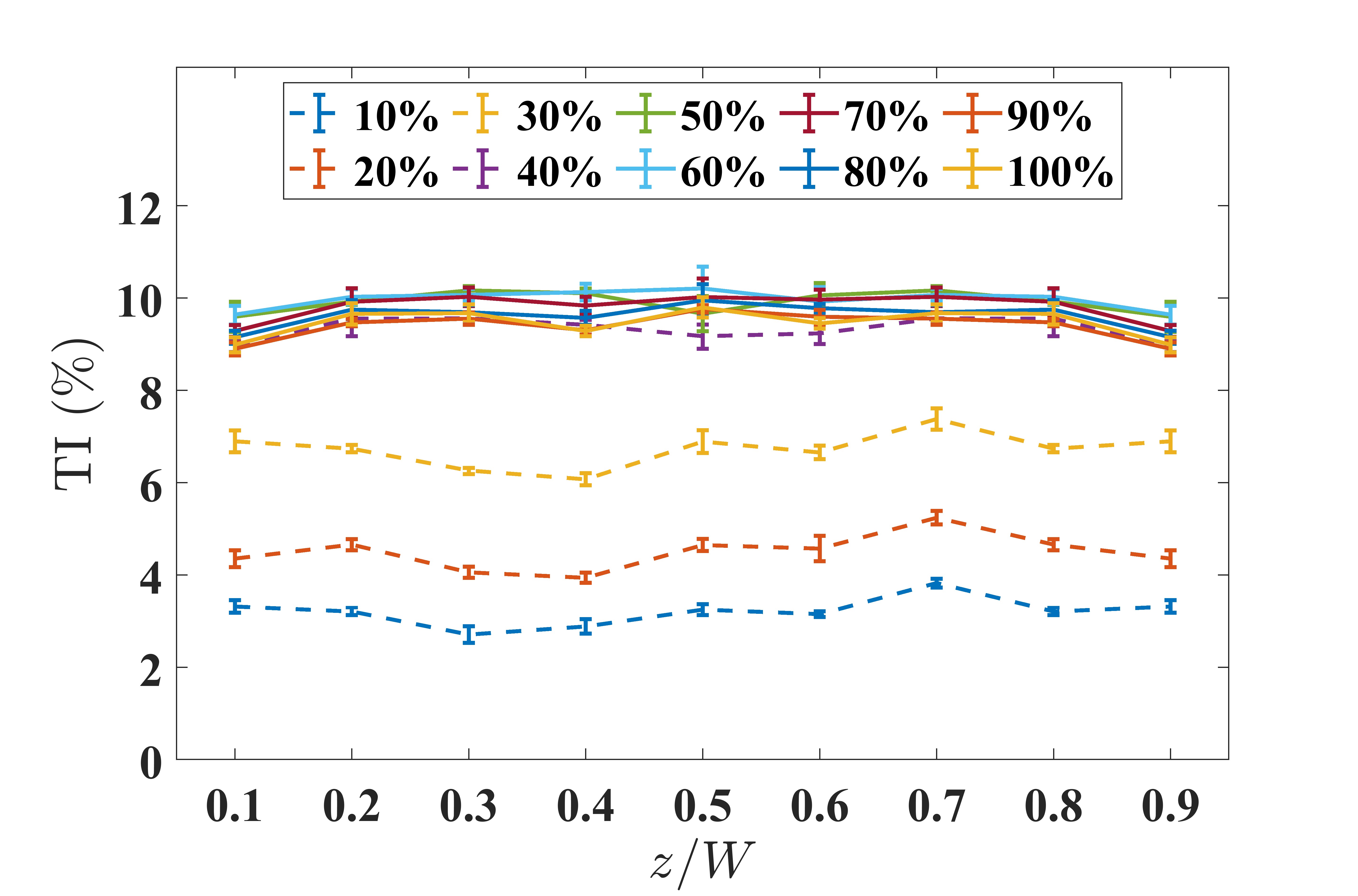}
\caption{ }
\label{fig:TIvsz}
\end{subfigure}
\caption{ Variation of turbulence intensity (TI) along the (a) non-dimensional height ($y/H$), and (b) non-dimensional width ($z/W$) as a function of PDM duty-cycle for the cross-section located at $x/L = 0.52$.}
\label{TIvsyandz}
\end{figure}

To investigate the turbulence flow field, the turbulent fluctuations $u^{'}$ are estimated to determine the TI of the flow field presented in Figure \ref{Umeanvsyandz}. Figures \ref{fig:tivsy} and \ref{fig:TIvsz} present the variation of the TI in the streamwise velocity component over a range of fan duty-cycles along the height and width of the test-section, respectively, for the cross-section located at a non-dimensional distance, $x/L$ = 0.52, from the fan-array, along the length of the test section. From the figures, it can be observed that the TI ranges between 2\% to 11\% across the height and width. TI exhibits an S-curve relationship with the fan duty-cycle, wherein TI increases with duty-cycles between 10\% to 40\% and nearly gets stabilized between 9\% - 11\% for duty-cycles above 40\%. The observations in Figure \ref{TIvsyandz} further extend the observations reported by Li et al. \cite{li2024aerodynamic}, where TI values were investigated across a test-section for 100\% and 50\% duty-cycles, and similar TI values were observed for both 100\% and 50\%. The maximum variation in TI is about 1.55\% and 0.72\% along the height and width, respectively, with a maximum standard deviation of approximately 0.25\% being observed. 

\begin{landscape}
\begin{table}[]
\captionsetup{justification=centering}
\caption{Comparison of the flow characteristics for various FAWT configurations under uniform flow conditions with outdoor measurements by Walpen et al. \cite{walpen2023real}}
\centering
\begin{tabular}{|p{4cm}|p{3cm}|p{3cm}|p{3cm}|p{1.5cm}|p{1.5cm}|p{1.5cm}|p{1.5cm}|}
\toprule
Flow characteristics in uniform flow conditions $\rightarrow$                     & \multicolumn{1}{c|}{}      & \multicolumn{1}{l|}{}   & \multicolumn{1}{l|}{}   & \multicolumn{1}{l|}{} & \multicolumn{1}{l|}{} & \multicolumn{1}{l|}{} &                                  \\ \midrule
FAWT $\downarrow$                                                            & Fan array size          & Flow Modulating Devices & Confined or Unconfined (test section) & $\overline{U_0}$ (m/s) & TI (\%)               & $\bar{\tau}$ (sec)         & $l_s$ (cm) \\ \midrule
Current work                                       & 10   x 10                  & No                      & Confined                & 1.6   - 7.6           & 1.63 - 18.35          & 30.68-62.68           & 2.17-3.22                        \\ \midrule
Walpen (2023)                                                        & 18   x 9 x 2 / 36 x 24 x 2 & No                      & Unconfined              & 16                    & 5 - 40                & --                    & 2   - 70                         \\ \midrule
Li   (2023)                                                          & 10   x 10                  & No                      & Unconfined              & 4.43-12.24            &   $\le$ 10                & --                   & --                              \\ \midrule
Liu   (2025)                                                         & 40 x 40                    & No                      & Unconfined              & 12.43-17.56           & 3 - 20                & --                    & --                               \\ \midrule
Chaudhary   (2025)                                                   & 2 x 5                      & Yes                     & Confined                & 3.38-6.65             & --                    & --                    & --                               \\ \midrule
\begin{tabular}[c]{@{}c@{}}Outdoor\\    \\ Measurements\end{tabular} & --                         & No                      & Unconfined              & 6.67                  & 32                    & --                    & 55.7   m                         \\ \bottomrule
\end{tabular}
\label{Table:ComparisonofFAWT}
\end{table}
\end{landscape}

\begin{figure}
\centering
\begin{subfigure}[h]{0.43\textwidth}
\centering
\includegraphics[width=\textwidth,trim={0cm 0cm 2cm 0cm},clip]{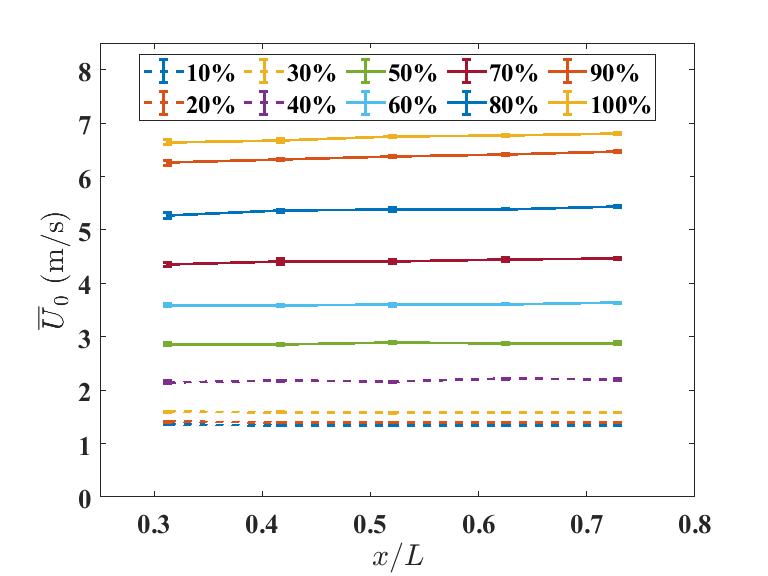}
\caption{ }
\label{fig:umeanvsx}
\end{subfigure}
\hfill
\begin{subfigure}[h]{0.49\textwidth}
\centering
\includegraphics[width=\textwidth,trim={4.5cm 0.6cm 9cm 0.4cm},clip]{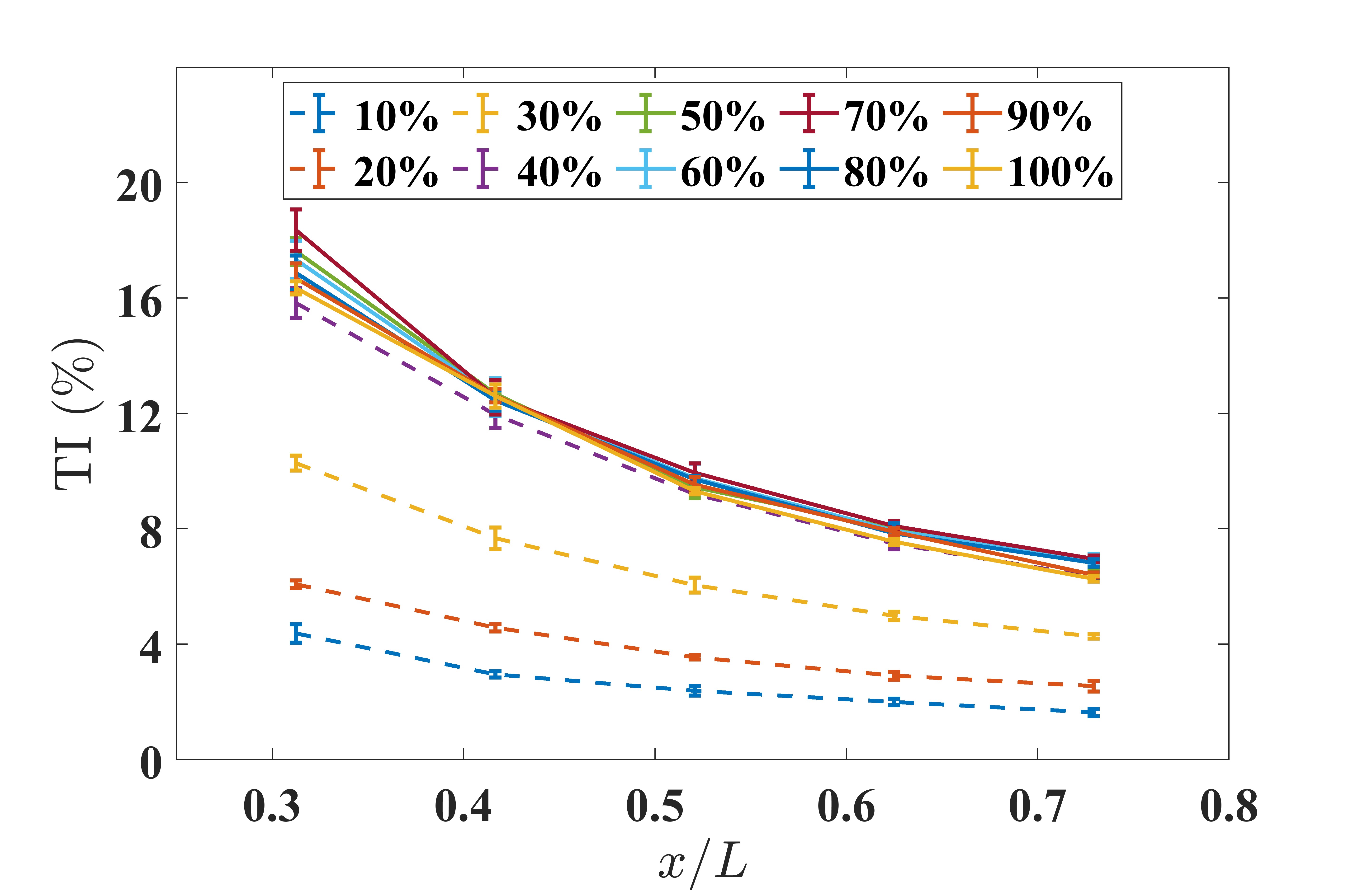}
\caption{ }
\label{fig:TIvsx}
\end{subfigure}
\caption{Variation of (a) mean velocity ($\overline{U_0}$) and (b) turbulent intensity (TI) measured at the center of each cross-section along the length of the test-section ($x/L$) as a function of PDM duty-cycle.}
\label{UmeanandTIvsx}
\end{figure}

To further understand how the flow field is evolving along the length of the tunnel, Figure \ref{fig:umeanvsx} and \ref{fig:TIvsx} present the $\overline{U_0}$ and TI respectively at the center of the test-section, i.e. $y/H$ = 0.5 and $z/W$ = 0.5, for five different $x/L$ = [0.31, 0.41,  0.52, 0.62, 0.73] as a function of fan duty-cycle. From Figure \ref{fig:umeanvsx}, it can be noticed that the wind speeds remain almost constant along the length of the test-section for a given duty-cycle. Figure \ref{fig:TIvsx} shows that the TI is maximum closer to the fan-array, and it reduces as we go away from the fan-array for all the fan duty-cycles ranging from 10\% - 100\%. This observation extends the observations presented in \cite{dougherty2018fan} for 100\% duty-cycle, Li et al. \cite{li2024aerodynamic} for 100\%, 80\% and 50\% duty-cycles and it can also be observed by Walpen et al. \cite{walpen2023real} where the TI is around 40\% at the exit of the fans which settles down and reaches to a 5\% intensity at a downstream distance of 1m from the fan array, with all fans operating uniformly at 25\% duty cycles and with a mean velocity of 5 m/s. The higher turbulent fluctuations have been attributed by Walpen et al. \cite{walpen2023real} to the sensor’s proximity with the fan blades and hub wakes. \ref{fig:TIvsx} also shows that the TI values are significantly lower for duty-cycles less than 40\%, even for $x/L$ = 0.31, and this observation further supports the observations made in Figures \ref{TIvsyandz}. The decrease in the turbulent fluctuations as $x/L$ increases signifies the transfer of turbulent kinetic energy from the free-stream to the turbulence boundary layer. Figures \ref{Umeanvsyandz} to \ref{UmeanandTIvsx} present the variation of the average freestream velocities ($\overline{U_0}$) and the turbulent fluctuations (TI) across the transverse, longitudinal, and cross-stream directions developed in the FAWT under uniform flow conditions with $\overline{U_0}$ $\in$ [1.6 $-$ 7.6] m/s and the TI $\in$ [1.63 $-$ 18.35] \%.  These values show some 
resemblance to the real-time atmospheric flow conditions measured by Walpen et al. \cite{walpen2023real} as shown in Table \ref{Table:ComparisonofFAWT} at the rooftop of a laboratory over two days by using an HWA. They recorded a lowest average airspeed of 1.66 m/s on a calm day and a velocity of 5.66 m/s on a windy day with streams of gusts flowing at 15 m/s. Despite the mean speed being higher on the windy day, the turbulent fluctuations were almost similar for both days, at approximately 29\%. Table \ref{Table:ComparisonofFAWT} compares the flow characteristics under uniform operating conditions for various FAWTs. It can be observed that $\overline{U_0}$ and TI observed in this work are along similar lines with other FAWT configurations.

\subsection{Scales of turbulence}
The turbulence scales describe the behaviour of the turbulent flow field. It identifies the largest and smallest turbulent structures that exist in the flow field, the energy mechanism that binds them, facilitating the interplay between the viscous and inertial forces, and their effect on the dynamics of the flow. The large-scale motions are mainly influenced by the geometry of the flow, whereas the small-scale motion strongly depends on the rate at which they receive energy from the dissipation of the large scales. The integral length scales represent the averaged scale of turbulent fluctuations. According to Taylor’s frozen turbulence Hypothesis, the integral length scales can be given by Camp et al. \cite{camp1995turbulence}.

\begin{equation}
\begin{aligned}
&l_s=\overline{U_0} \bar{\tau}, { where }\ \bar{\tau}=\int_0^\tau A C F(\tau) d \tau \text { and }\\
&A C F(\tau)=\frac{\frac{1}{\Delta T} \int_0^{\Delta T} u(t) u(t+\tau) d t}{\frac{1}{\Delta T} \int_0^{\Delta T} u^2(t) d t}
\end{aligned}
\end{equation}

where $l_s$ is the integral length scale, and $\bar{\tau}$ denotes the integral time scale, ACF represents autocorrelation, $u(t)$ is the instantaneous fluid streamwise velocity component at a time $t$, $\Delta$T represents the time interval of the experiment, which is 10 seconds for the current study, and $\tau$ represents the time lag in seconds. The mathematical expression in Eq.1 for $l_s$ would represent the product of the mean freestream velocity with the physical area under the ACF($\tau$) curve from 0 to T, which would represent the time lag at which ACF($\tau$) first becomes zero.

\begin{table}[h]
\centering
\captionsetup{justification=centering}
\caption{Time and Length scales variation along $x/L = \{0.31, 0.41, 0.52, 0.62, 0.72\}$ with $y/H = z/W = 0.5$ for 100\% duty cycle}
\begin{tabular}{|l|l|c|c|c|}
\hline$x / L$ & $\bar{\tau}(\mathrm{sec})$ & $\sigma_\tau(\mathrm{sec})$ & $l_s(\mathrm{~cm})$ & $\sigma_{l_s}(\mathrm{~cm})$ \\
\hline 0.31 & 30.682 & 1.540 & 2.246 & 0.117 \\
\hline 0.41 & 31.374 & 0.575 & 2.321 & 0.042 \\
\hline 0.52 & 34.511 & 2.453 & 2.558 & 0.168 \\
\hline 0.62 & 36.014 & 3.563 & 2.681 & 0.265 \\
\hline 0.72 & 38.211 & 2.085 & 2.858 & 0.155 \\
\hline
\end{tabular}
\label{table:timelengthscalesalongx}
\end{table}

\begin{table}[h]
\centering
\captionsetup{justification=centering}
\caption{Time and Length scales variation at duty-cycles $\in \{60\%, 80\%, 100\%\}$ for $y/H = z/W = 0.5$, and $x/L=0.52$.}
\begin{tabular}{|c|c|c|c|c|}
\hline duty-cycle $(\%)$ & $\bar{\tau}(\mathrm{sec})$ & $\sigma_\tau(\mathrm{sec})$ & $l_s(\mathrm{~cm})$ & $\sigma_{l_s}(\mathrm{~cm})$ \\
\hline 60 & 62.683 & 2.170 & 2.481 & 0.093 \\
\hline 80 & 42.520 & 2.222 & 2.512 & 0.139 \\
\hline 100 & 34.511 & 2.453 & 2.558 & 0.168 \\
\hline
\end{tabular}
\label{table:timelengthscalesfordc}
\end{table}

\begin{table}[h]
\centering
\captionsetup{justification=centering}
\caption{Time and Length scales variation along $y/H = [0.1 - 0.9]$ with $x/L =0.52$ and $z/W = 0.5$ for 100\% duty cycle}
\begin{tabular}{|l|l|c|c|c|}
\hline$y / H$ & $\bar{\tau}(\mathrm{sec})$ & $\sigma_\tau(\mathrm{sec})$ & $l_s(\mathrm{~cm})$ & $\sigma_{l_s}(\mathrm{~cm})$ \\
\hline 0.1 & 36.944 & 0.587 & 2.536 & 0.039 \\
\hline 0.2 & 44.233 & 5.148 & 3.221 & 0.355 \\
\hline 0.3 & 31.994 & 2.385 & 2.396 & 0.184 \\
\hline 0.4 & 32.734 & 2.398 & 2.449 & 0.170 \\
\hline 0.5 & 34.511 & 2.453 & 2.558 & 0.168 \\
\hline 0.6 & 34.842 & 1.796 & 2.484 & 0.209 \\
\hline 0.7 & 38.101 & 2.666 & 2.800 & 0.196 \\
\hline 0.8 & 44.233 & 5.148 & 3.221 & 0.35 \\
\hline 0.9 & 36.944 & 0.587 & 2.536 & 0.039 \\
\hline
\end{tabular}
\label{table:timelengthscalesalongy}
\end{table}

\begin{table}[h]
\centering
\captionsetup{justification=centering}
\caption{Time and Length scales variation along $z/W = [0.1 - 0.9]$ with $x/L =0.52$ and $y/H = 0.5$ for 100\% duty cycle.}
\begin{tabular}{|l|l|c|c|c|}
\hline$z / W$ & $\bar{\tau}(\mathrm{sec})$ & $\sigma_\tau(\mathrm{sec})$ & $l_s(\mathrm{~cm})$ & $\sigma_{l_s}(\mathrm{~cm})$ \\
\hline 0.1 & 36.538 & 1.327 & 2.699 & 0.103 \\
\hline 0.2 & 33.175 & 0.951 & 2.501 & 0.067 \\
\hline 0.3 & 40.947 & 2.234 & 2.850 & 0.156 \\
\hline 0.4 & 34.503 & 3.485 & 2.524 & 0.248 \\
\hline 0.5 & 34.511 & 2.453 & 2.558 & 0.168 \\
\hline 0.6 & 31.921 & 1.255 & 2.393 & 0.089 \\
\hline 0.7 & 32.641 & 0.889 & 2.494 & 0.055 \\
\hline 0.8 & 33.175 & 0.951 & 2.501 & 0.067 \\
\hline 0.9 & 36.538 & 1.327 & 2.699 & 0.103 \\
\hline
\end{tabular}
\label{table:timelengthscalesalongz}
\end{table}

\begin{figure}
\centering
\begin{subfigure}[b]{0.49\textwidth}
\centering
\includegraphics[width=\textwidth,trim={2cm 0.6cm 9cm 0.4cm},clip]{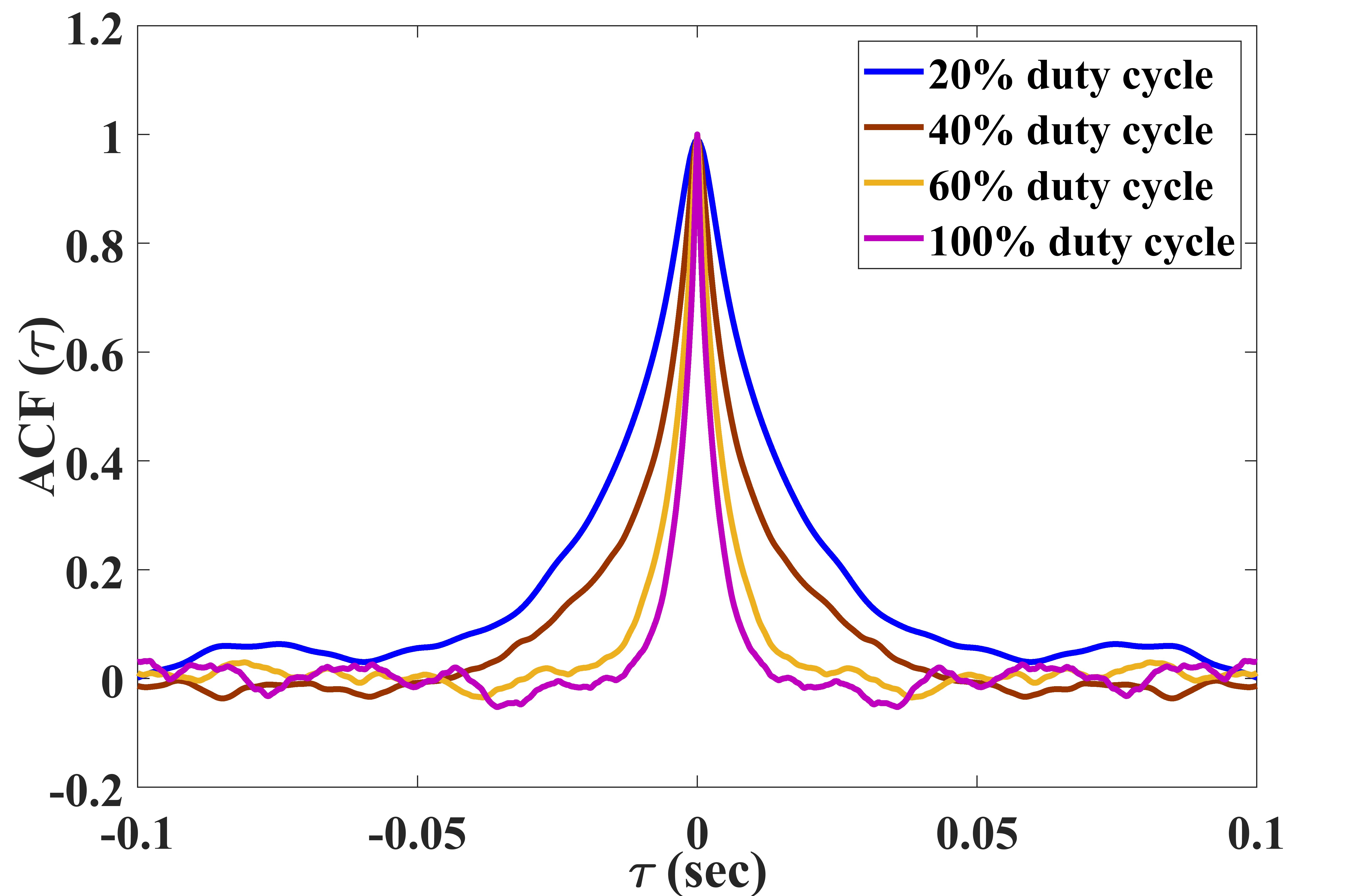}
\caption{ }
\label{fig:ACFcsDC}
\end{subfigure}
\hfill
\begin{subfigure}[b]{0.49\textwidth}
\centering
\includegraphics[width=\textwidth,trim={2cm 0.6cm 9cm 0.4cm},clip]{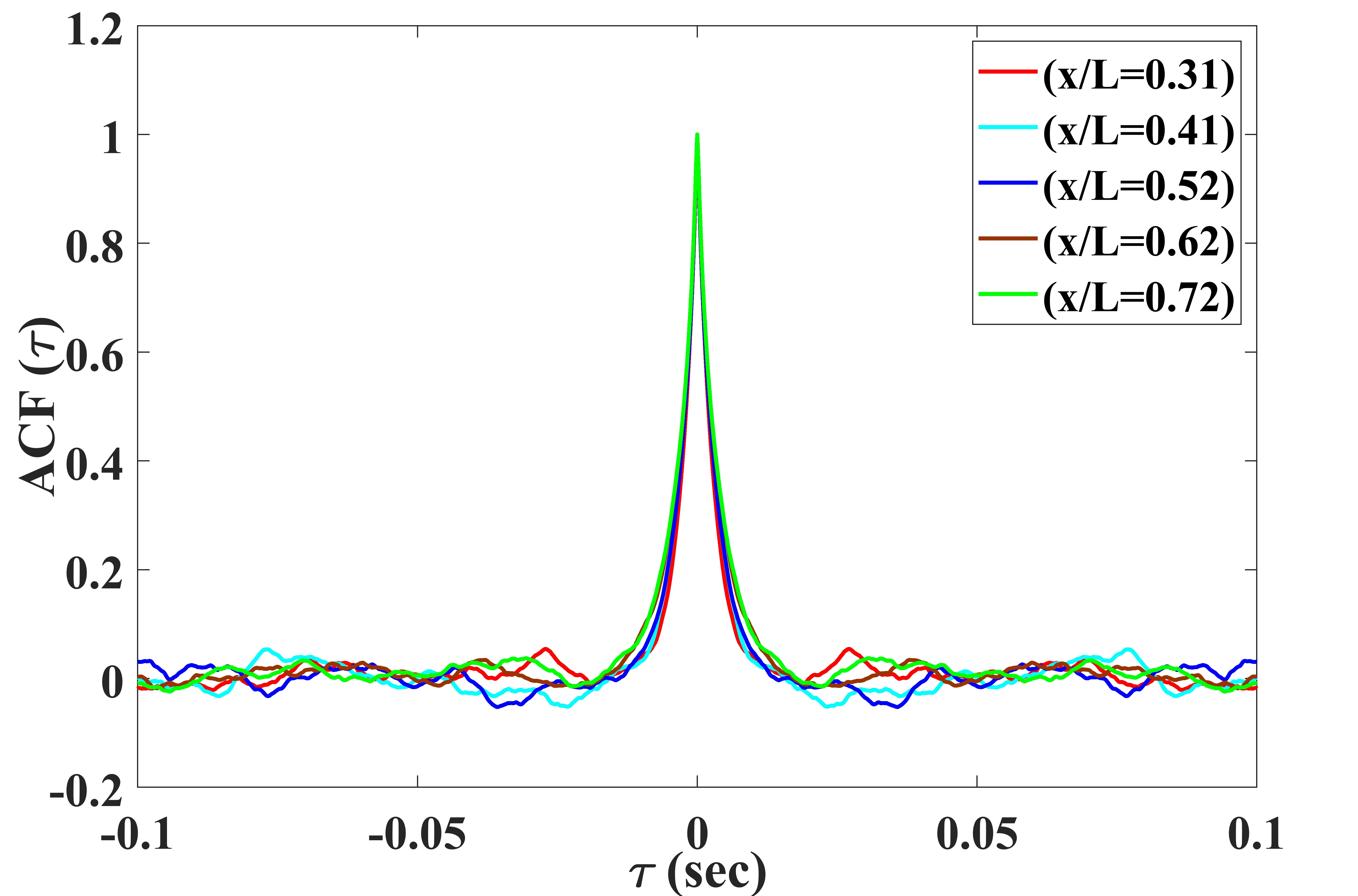}
\caption{ }
\label{fig:ACFvsL}
\end{subfigure}
\caption{ Comparison of the ACF (a) for duty-cycles $\in \{20\%, 40\%, 60\%, 100\%\}$ at $x/L = y/H = z/W = 0.5$ (b) and at various locations across the non-dimensional length ($x/L$) at $y/H$ and $z/W = 0.5$.}
\label{fig:ACFvsDCandL}
\end{figure}

\begin{figure}
\centering
\begin{subfigure}[b]{0.49\textwidth}
\centering
\includegraphics[width=\textwidth,trim={2cm 0.6cm 9cm 0.4cm},clip]{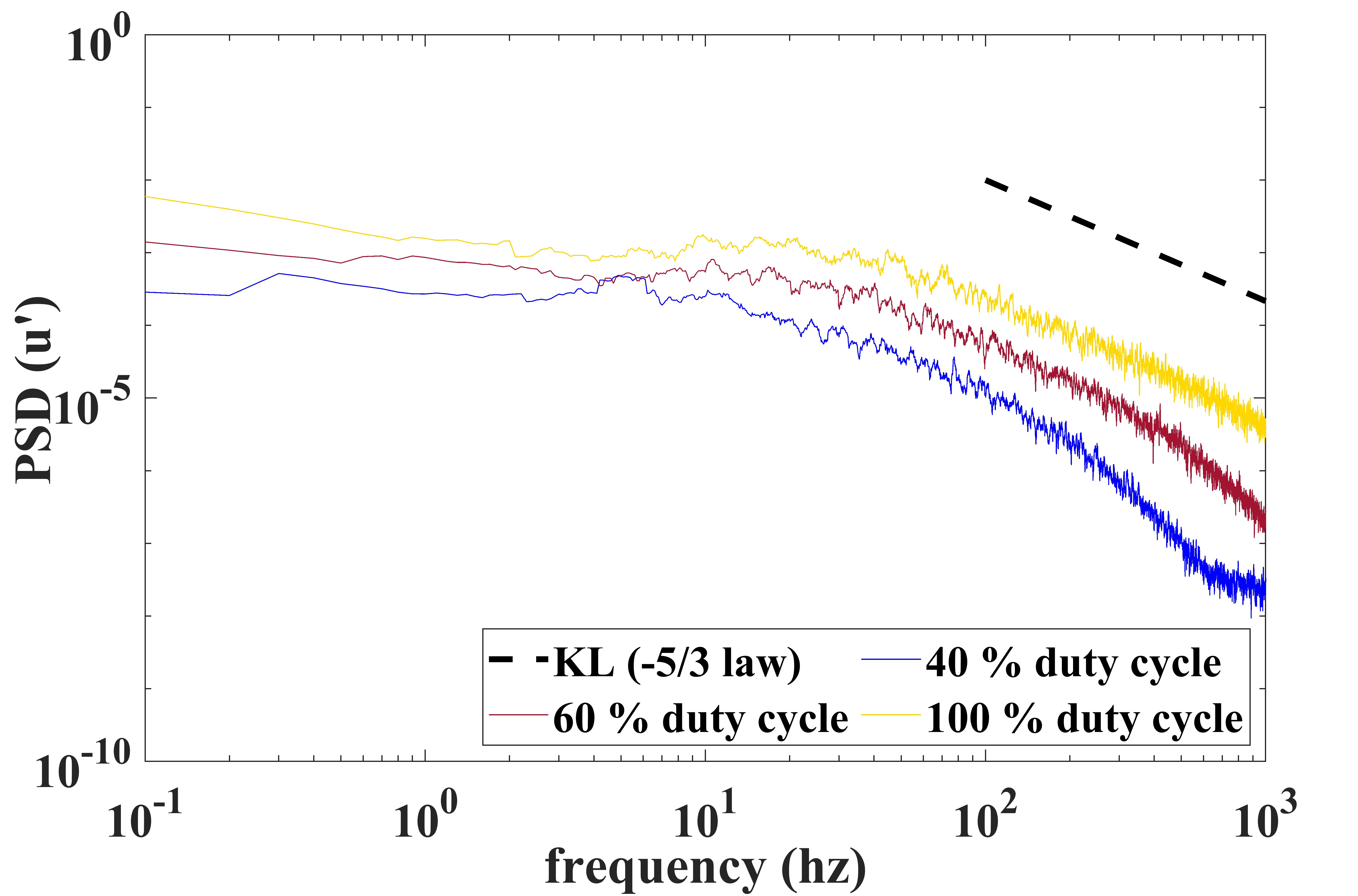}
\caption{ }
\label{fig:PSDcsDC}
\end{subfigure}
\hfill
\caption{PSD of TKE at the center of the test-section, i.e., $x/L = 0.5$, $y/H = 0.5$, and $z/W = 0.5$ for various duty-cycles. The dashed line in the figure is a representative straight line on the log scale that has a slope of -5/3.}
\end{figure}

Figure \ref{fig:ACFcsDC} presents the typical ACF plot for the streamwise velocity component at $x/L$ = 0.52, $y/H = 0.5$, and $z/W = 0.5$ as a function of fan duty-cycle for time lag $\tau \in [$-$0.1,0.1]$. From this figure, one can observe a broader and wider ACF curve for lower duty cycles of 20\%. As the duty cycle increases, the ACF plot becomes narrower and approaches zero correlation earlier when compared to their lower duty cycle counterparts. Hence, the area underneath the curve and the integral time scale ($\bar{\tau}$) reduce with the increasing duty cycle. Figure \ref{fig:ACFvsL} presents the variation of ACF along the length, $x/L = \{0.32,0.41,0.52,0.62,0.72\}$, of the test section for $y/H = 0.5$, $z/W = 0.5$, and 100\% duty-cycle. The ACF curve widens slightly as we move away from the fan array, thereby increasing the area under the ACF, representing an increase in the integral time scale of the eddy along the length of the tunnel. A detailed discussion on the physical phenomenon behind the increase in the integral time scale of the eddy is presented in the next paragraph.

Table \ref{table:timelengthscalesalongx} summarizes the evolution of the integral time and length scales in the streamwise direction for the 100\% duty-cycle operating condition at $y/H$ = 0.5 and $z/W$ = 0.5. It can be observed from the table that as we move farther downstream from the fan array, $l_s$ also increases and thereby increasing the size of $l_s$. In Figure \ref{UmeanandTIvsx}, it has been shown that the TI decreases as $x/L$ increases, which in turn would represent the reduction of turbulent kinetic energy. Typically, it is expected that the reduction in turbulent kinetic energy is associated with the dissipation of energy from larger eddies into smaller eddies, thereby potentially reducing the $l_s$. However, this physical phenomenon contradicts the data presented in Table \ref{table:timelengthscalesalongx}. In this work, the increase in the $l_s$ can be attributed to the energy that is being injected into the eddy from the freestream wind shear generated by the fan array, which can result in stretching of the eddy in the streamwise direction. Similar observations were reported by Walpen et al. \cite{walpen2023real}, wherein they observed a 2 cm integral length scale at 0.1 m downstream, rising to almost 30 to 70 cm at 5 cm downstream distance, under uniform and steady conditions with all fans operating at 25\% duty cycle. For a constant $\overline{U_0}$, the integral length scale is directly proportional to the integral time scale according to Taylor’s frozen turbulence hypothesis, which explains the growing integral time scales ($\bar{\tau}$)as the sensor’s position is moved further downstream along the length of the tunnel. In Table \ref{table:timelengthscalesalongx}, the columns $\sigma_{\tau}$  and $\sigma_{ls}$ would represent the standard deviation in $\bar{\tau}$ and $l_s$ respectively.

Table \ref{table:timelengthscalesfordc} presents the $\bar{\tau}$ and $l_s$ determined across various duty-cycles at the center of the test-section, i.e., $x/L$ = 0.52 and $y/H$ = $z/W$ = 0.5. It can be observed from the table that the $\bar{\tau}$ gradually decreases with increasing fan duty cycle and is in good agreement with the observations made in Figure \ref{fig:ACFvsDCandL}. However, it should be noted that $l_s$ remains approximately the same across the duty cycles. The maximum variation in the mean $l_s$  across the duty-cycles is 0.77mm. Tables \ref{table:timelengthscalesalongy} and \ref{table:timelengthscalesalongz} report the $\bar{\tau}$ and $l_s$   calculated along the height and width of the test-section for the cross-sectional plane located at $x/L$ = 0.52. One can observe that the $\bar{\tau}$ and $l_s$ remain approximately the same across the cross-section.

Table \ref{Table:ComparisonofFAWT} showcases the comparison between the aerodynamic characteristics achieved by different fan-array structures developed in recent years. When compared against atmospheric and outdoor conditions measured by Walpen et al. \cite{walpen2023real}, this data reflects that these fan array tunnels are effective at generating wind environments that closely mirror real-world outdoor conditions. This has direct implications for aerodynamic applications, proving valuable for tasks like evaluating the flight dynamics of unmanned vehicles or studying the aerodynamic forces acting on rigid or flexible airborne structures.

\subsection{Power Spectral Densities (PSD)}

\begin{figure}
\centering
\begin{subfigure}[b]{0.49\textwidth}
\centering
\includegraphics[width=\textwidth,trim={3cm 0.6cm 9cm 0.4cm},clip]{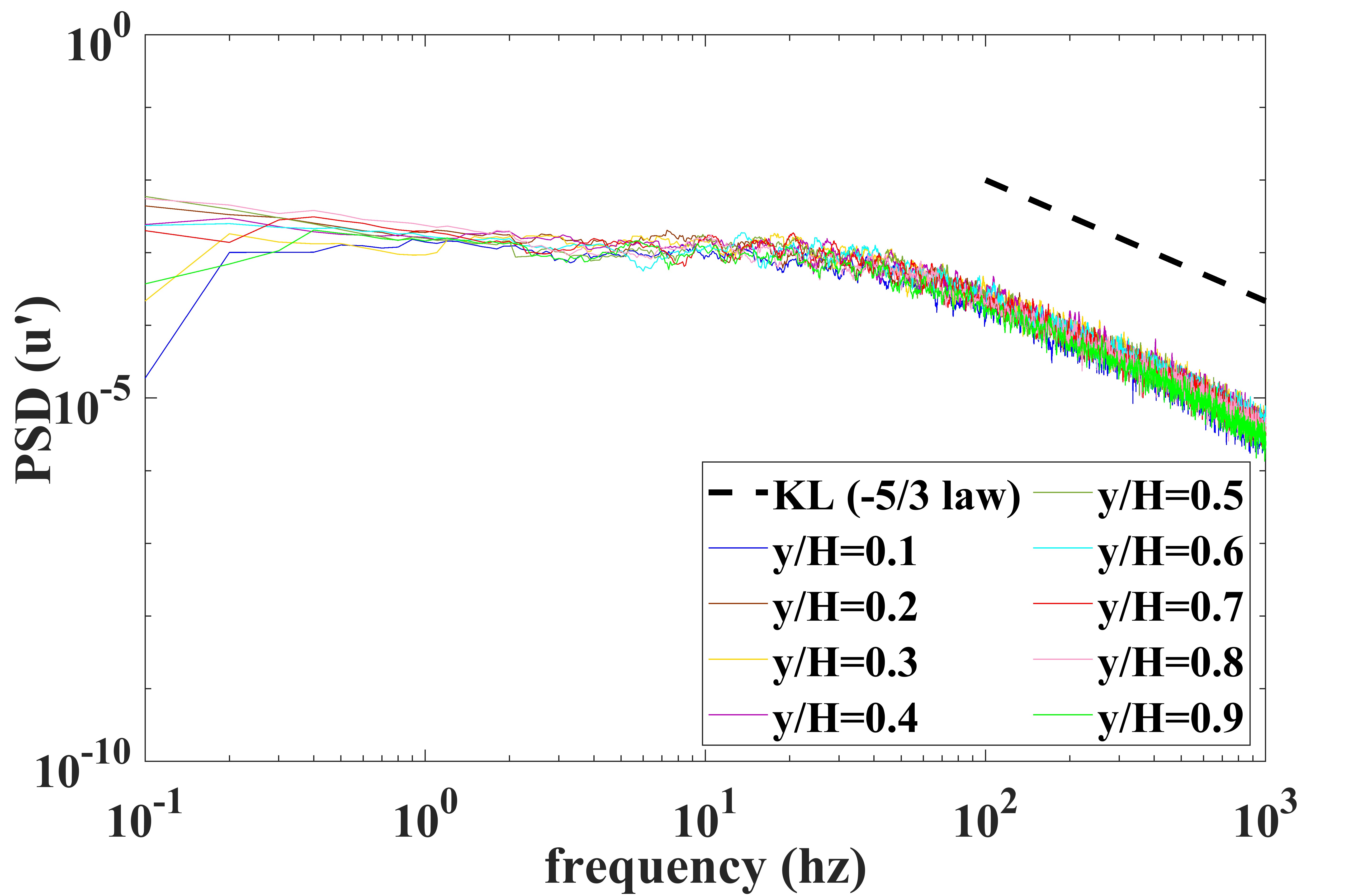}
\caption{ }
\label{fig:PSDvsy}
\end{subfigure}
\hfill
\begin{subfigure}[b]{0.49\textwidth}
\centering
\includegraphics[width=\textwidth,trim={2cm 0.6cm 9cm 0.4cm},clip]{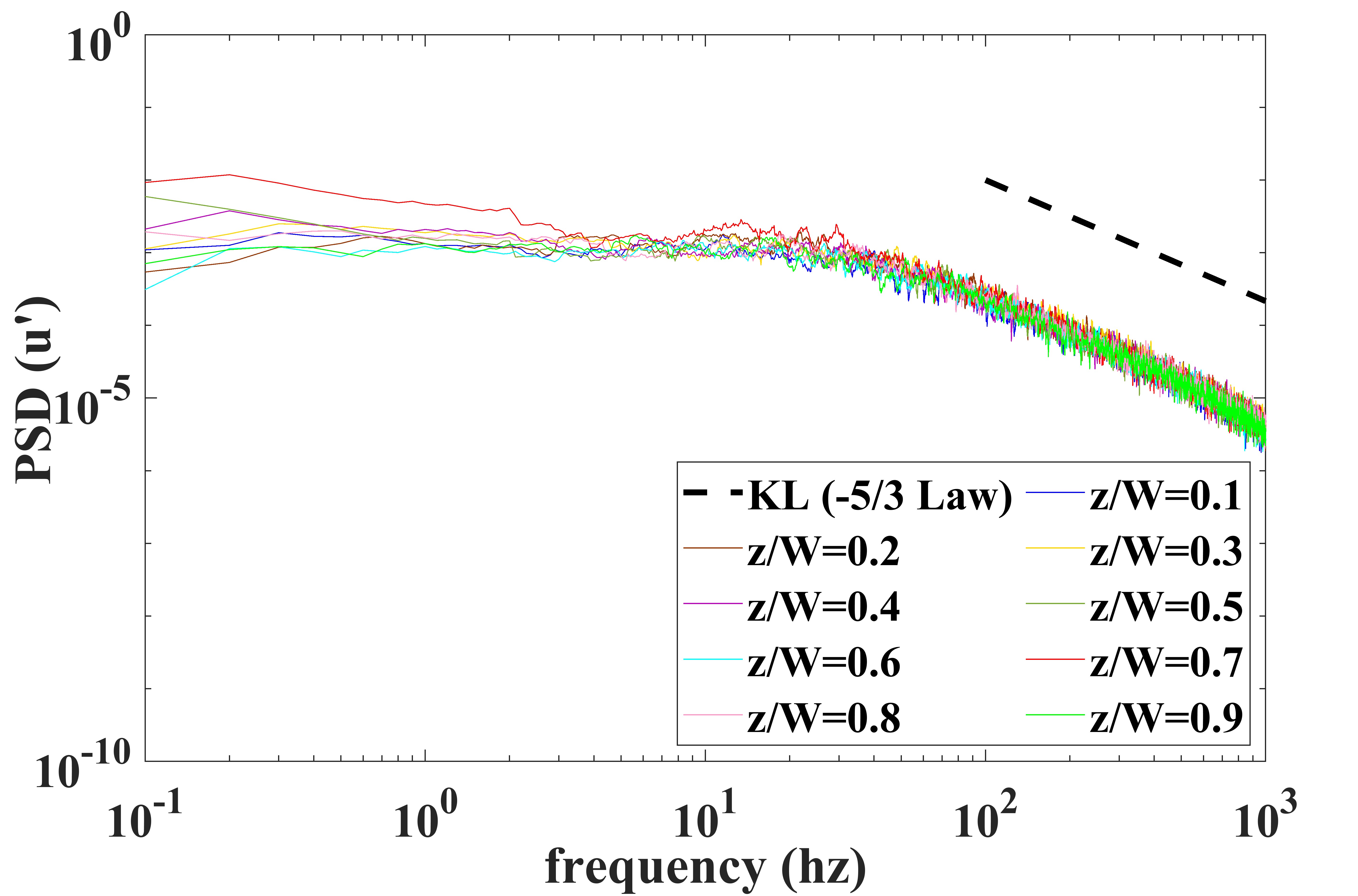}
\caption{ }
\label{fig:PSDvsz}
\end{subfigure}
\begin{subfigure}[b]{0.49\textwidth}
\centering
\includegraphics[width=\textwidth,trim={2cm 0.6cm 9cm 0.4cm},clip]{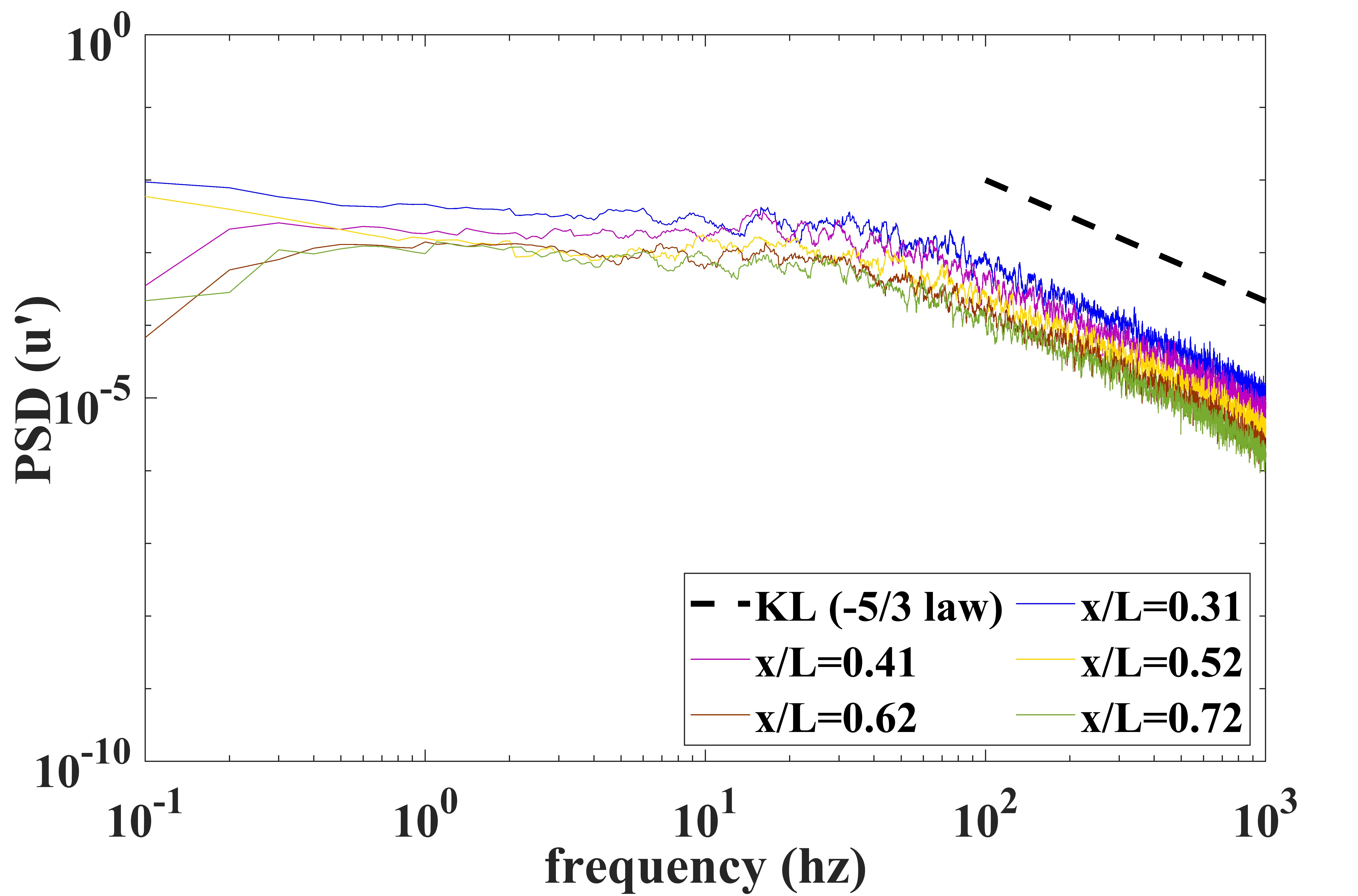}
\caption{ }
\label{fig:PSDvsx}
\end{subfigure}
\caption{ Variation of the TKE’s PSD at for the 100\% duty cycle (a) along the non-dimensional height ($y/H$) for $x/L = 0.52$ and $z/W = 0.5$, (b) along the non-dimensional width ($z/W$) for $x/L = 0.52$ and $y/H = 0.5$, and (c) along the non-dimensional length ($x/L$) for $y/H = 0.5$ and $z/W = 0.5$. The dashed line in the figure is a representative straight line on the log scale that has a slope of -5/3.}
\label{fig:PSDvsyzandx}
\end{figure}

 The power spectral density (PSD) of the turbulent kinetic energy (TKE) characterizes the energy transfer mechanism in the inertial length scale regime. The magnitude of the PSD for a given frequency would also signify the strength of the turbulent fluctuations/intensities. Figure \ref{fig:PSDcsDC} represents the PSD plot of the TKE for a range of duty cycles for the point at the center of the test section, i.e., $x/L$ = 0.5, $y/H$ = 0.5, and $z/W$ = 0.5. The flat part of the spectra, as can be seen from figures \ref{fig:PSDvsyzandx}, from 0-100 Hz, can be attributed to the turbulence generated by the fans (\cite{walpen2023real}), which slowly dissipates as it moves further downstream along the tunnel. The dashed line in the figure is a representative straight line in the log scale, having a slope of -5/3, and one can observe from the figure that the curves are parallel to the -5/3 slope line for the frequencies between 100 to 1000 Hz. The magnitude PSD at a given frequency clearly shows that the TKE reduces with decreasing duty cycles, and this observation conforms with the TI analysis presented in Section 3.1. From the figure, it can be observed that the Kolmogorov time scales for the FAWT at the 40\% duty cycle are bigger than their counterparts from higher duty cycles. This observation is similar to the observation made for $\bar{\tau}$ in Table \ref{table:timelengthscalesfordc}. Figure \ref{fig:PSDvsyzandx} presents the variation of PSD in TKE along the height, width, and length of the test section. It can be observed that the PSD curves are parallel to the -5/3 slope representative straight line. One can observe from Figures \ref{fig:PSDvsy} and \ref{fig:PSDvsz} that the PSD curves for different points on the same cross-section overlap. Figure \ref{fig:PSDvsx} shows that the TKE for a given frequency decreases as we move away from the fan array, signifying the reduction in the TI with increasing $x/L$ (a similar observation has been made in Section 3.1).
 
\subsection{Effect of fan geometry on turbulence measurements}

\begin{figure}
\centering
\includegraphics[width=\textwidth,trim={0.3cm 2cm 0cm 2.5cm},clip]{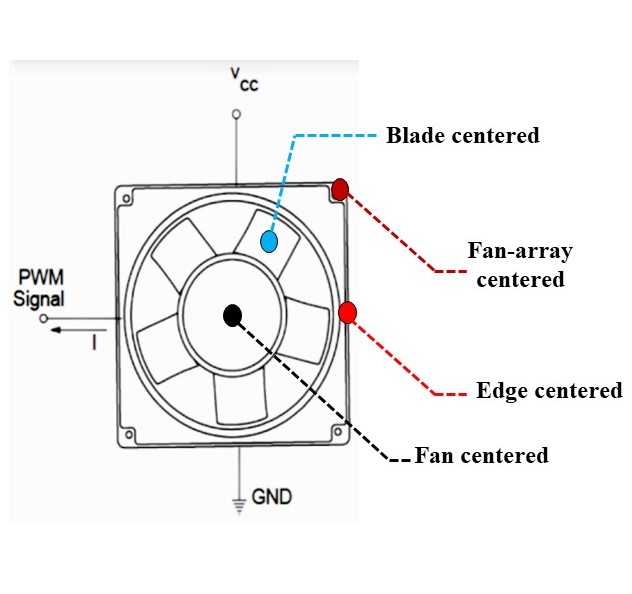}
\caption{Schematic representation of four distinct regions of a BLDC fan}
\label{fig:fancentres}
\end{figure}

\begin{figure}
\centering
\begin{subfigure}[b]{0.49\textwidth}
\centering
\includegraphics[width=\textwidth,trim={3cm 0.6cm 9cm 0.4cm},clip]{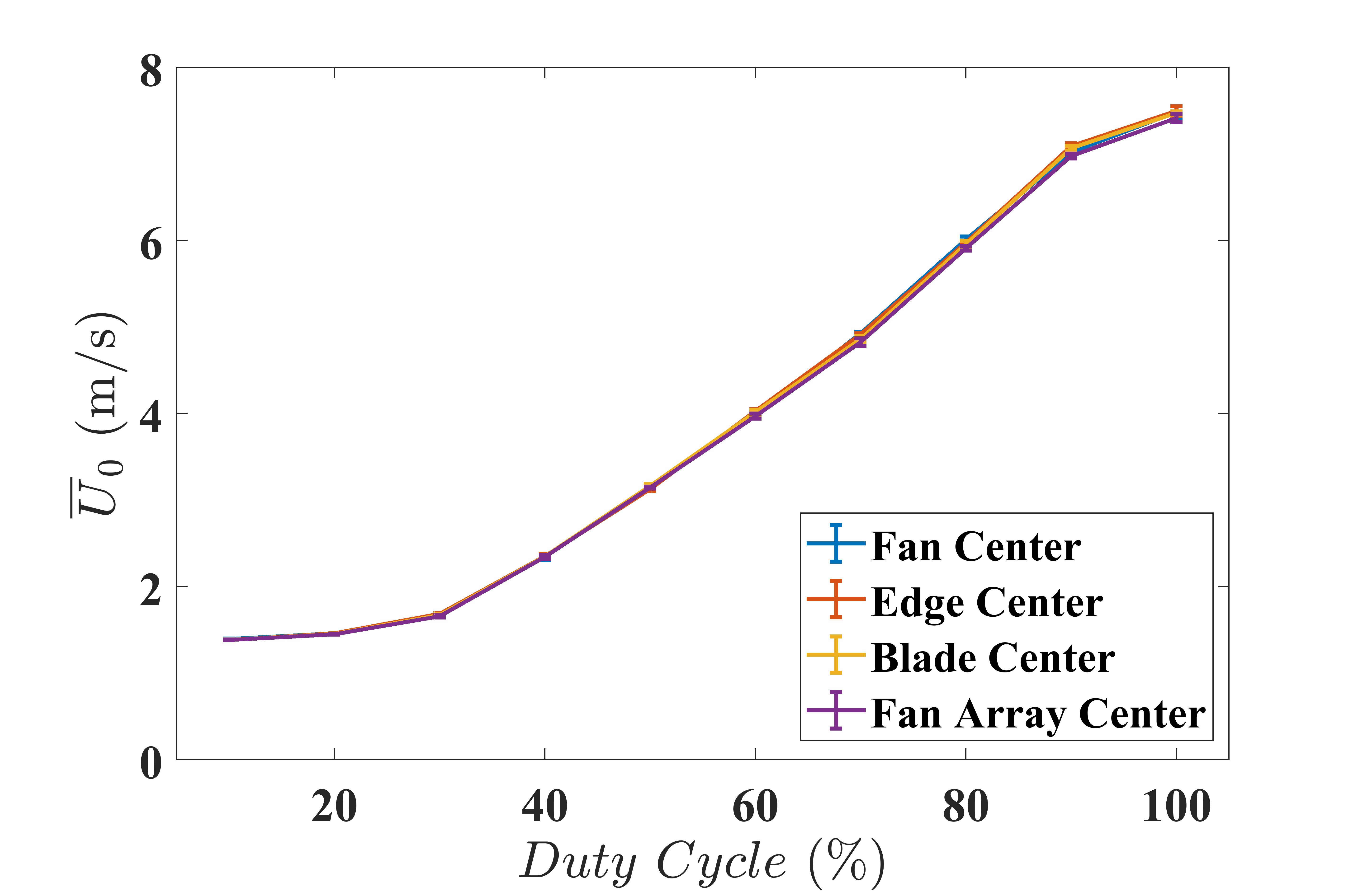}
\caption{ }
\label{fig:Umeanvscenters}
\end{subfigure}
\hfill
\begin{subfigure}[b]{0.49\textwidth}
\centering
\includegraphics[width=\textwidth,trim={3cm 0.6cm 9cm 0.4cm},clip]{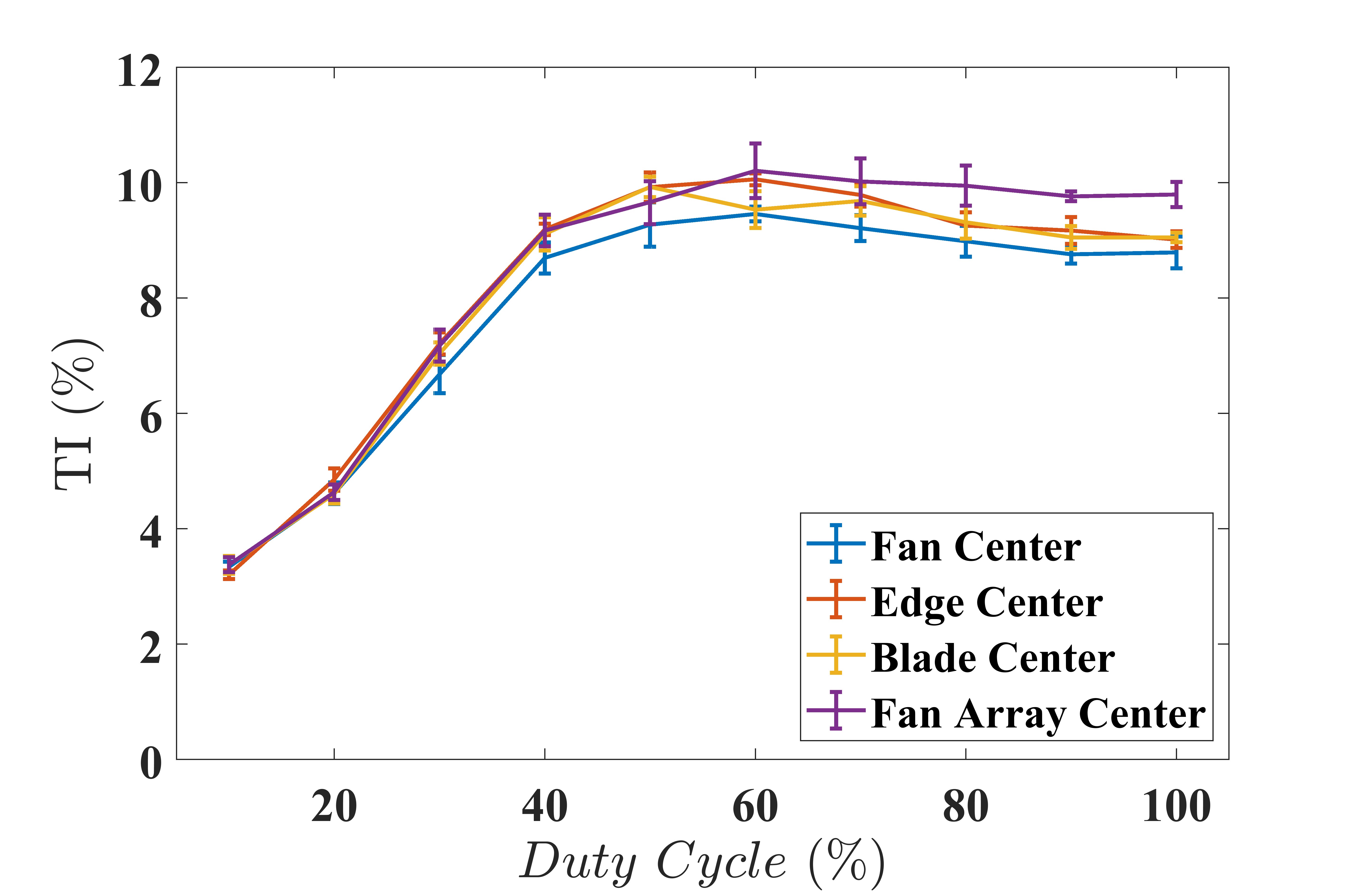}
\caption{ }
\label{fig:TIvscentres}
\end{subfigure}
\caption{ Variation (a) TI, and (b) Mean velocity distribution ($\overline{U_0}$) at the 4 distinct regions of the fan $\in$ {FC, EC, BC, FAC} for the cross-section $x/L = 0.52$.}
\label{fig:UmeanandTIvscenters}
\end{figure}

\begin{figure}
\centering
\begin{subfigure}[b]{0.49\textwidth}
\centering
\includegraphics[width=\textwidth,trim={3cm 0.6cm 9cm 0.4cm},clip]{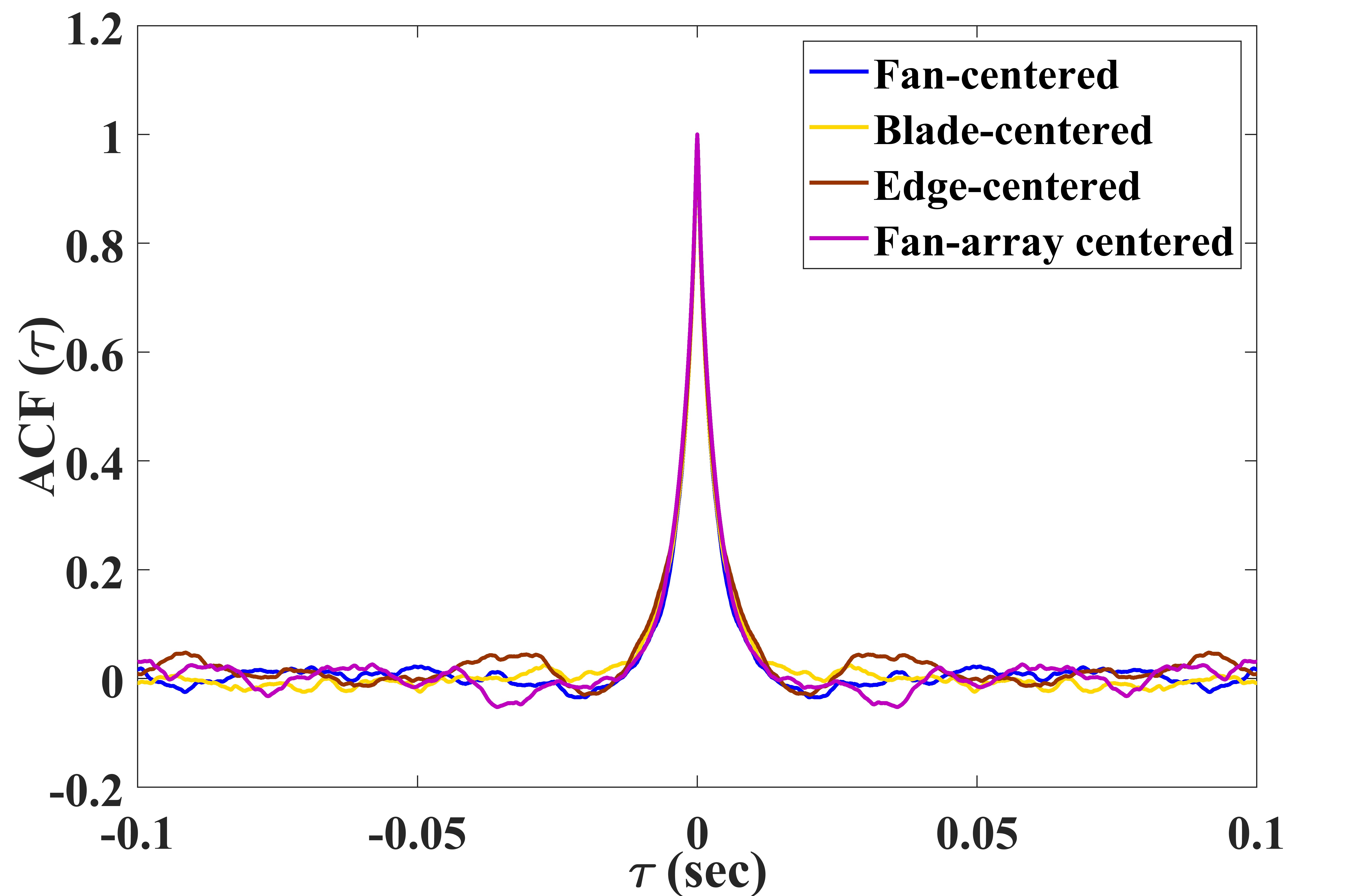}
\caption{ }
\label{fig:ACFvscentres}
\end{subfigure}
\hfill
\begin{subfigure}[b]{0.49\textwidth}
\centering
\includegraphics[width=\textwidth,trim={2cm 0.6cm 9cm 0.4cm},clip]{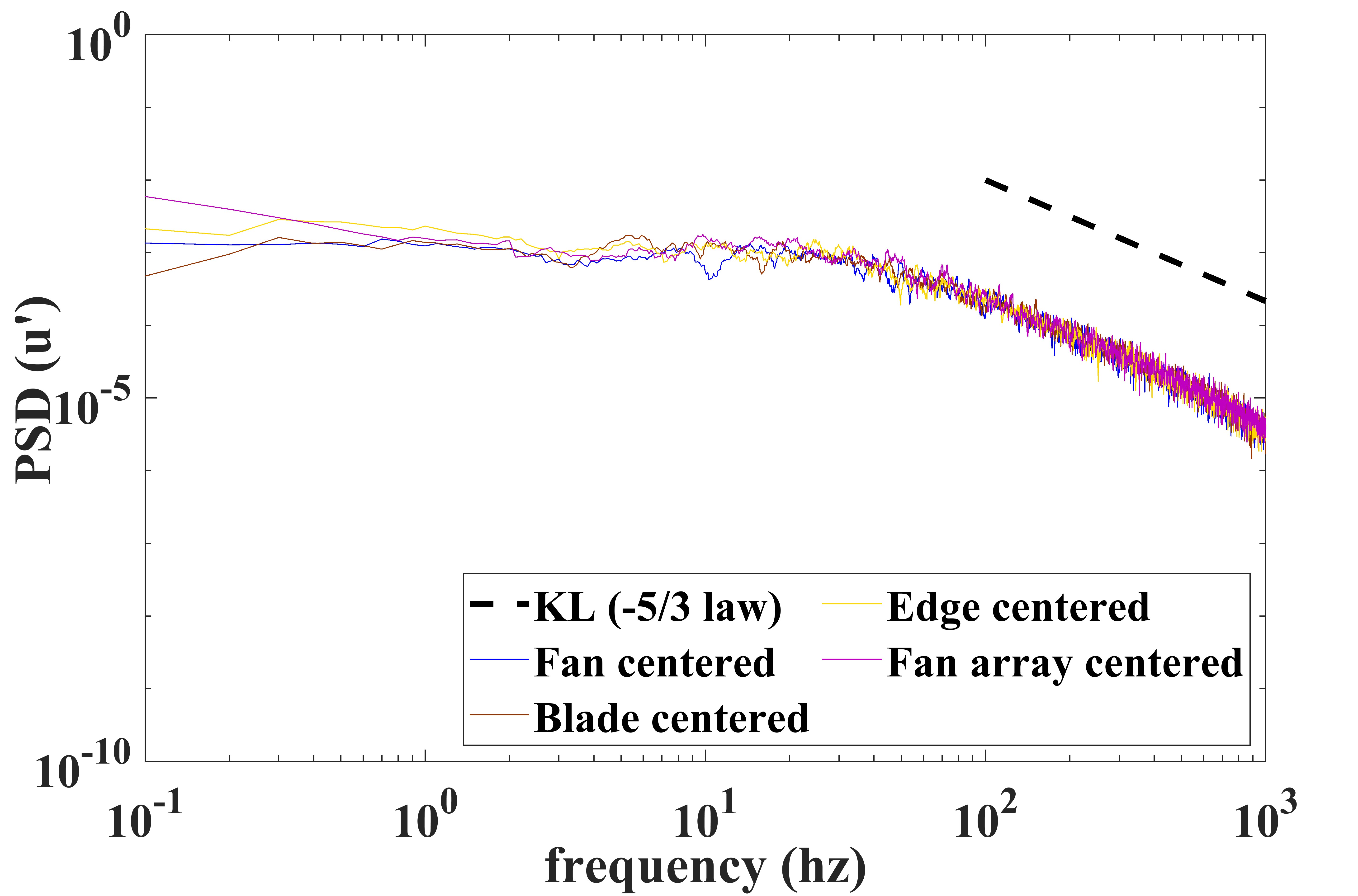}
\caption{ }
\label{fig:PSDvscentres}
\end{subfigure}
\caption{Comparison of (a) ACF and (b) PSD at the 4 distinct regions of the fan $\in$ {FC, EC, BC, FAC} for the cross-section x/L = 0.52.}
\label{fig:ACFandPSDvscenters}
\end{figure}

To understand the effect of fan geometry on the turbulence characteristics, Figure \ref{fig:fancentres} identifies four distinct regions of the fan: (a) fan/hub center (FC), (b) a point at the center of the fan blades i.e. the blade center (BC), (c) a point on the edge of the fan’s outer casing (EC), and (d) a point on the corner of the fan’s outer casing which would represent a 2×2 fan-array’s center (FAC). To understand the effect of fan geometry, the fan whose hub is located at $y/H$ = 0.55 and $z/W$ = 0.45 has been selected. Figure \ref{fig:Umeanvscenters} presents  ($\overline{U_0}$) measurements at $x/L$ = 0.52 for duty-cycles ranging from 10\%-100\%, and it can be observed that the effect of fan geometry is negligible on the $\overline{U_0}$. However, from Figure \ref{fig:TIvscentres}, the effect of fan geometry is evident on the TI for higher duty cycles. It can also be observed that the fan/hub center exhibits approximately 1\% lower TI when compared to the fan-array center (FAC). Figures \ref{fig:ACFvscentres} and \ref{fig:PSDvscentres} present the ACF and PSD for the locations FC, BC, EC, and FAC sites, respectively. Both the ACF and PSD curves mostly overlap for all the locations. To have a closer look into $\bar{\tau}$ and $l_s$, Table \ref{table:timelengthscalesalongframeofref} summarizes the integral time and length scales of the flow field for the 100\% duty cycle.

\begin{table}[h]
\centering
\captionsetup{justification=centering}
\caption{Time and Length scale variations at the 4 distinct regions of the fan $\in \{FC, EC, BC, FAC\}$ at $x/L = y/H = z/W = 0.5$.}
\begin{tabular}{|c|c|c|c|c|}
\hline Frame of reference & $\bar{\tau}(\mathrm{sec})$ & $\sigma_\tau(\mathrm{sec})$ & $l_s(\mathrm{~cm})$ & $\sigma_{l_s}(\mathrm{~cm})$ \\
\hline FC & 32.938 & 0.631 & 2.461 & 0.044 \\
\hline BC & 32.343 & 2.136 & 2.416 & 0.158 \\
\hline EC & 32.938 & 0.631 & 2.461 & 0.044 \\
\hline FAC & 34.511 & 2.453 & 2.558 & 0.168 \\
\hline
\end{tabular}
\label{table:timelengthscalesalongframeofref}
\end{table}

It can be observed that the $l_s$ for the FAC location is slightly bigger compared to other locations, and this observation is in line with the observations made in Figure \ref{fig:UmeanandTIvscenters}. In this section, the evolution of the turbulent time and length scales ($\bar{\tau}$ and ($l_s$) across the non-dimensional length, width, and height ($x/L$, $y/H$, $z/W$) of the FAWT, under various flow conditions, has been extensively discussed and presented. As can be seen from Tables \ref{table:timelengthscalesalongx} to \ref{table:timelengthscalesalongframeofref}, the integral length scales observed are in the range of  $l_s\in$ [2.17, 3.22] cm, and the integral time scales, $\bar{\tau} \in$ [30.68, 62.68] cm. 

\subsection{Pressure Measurements}

\begin{figure}
\centering
\begin{subfigure}[b]{0.49\textwidth}
\centering
\includegraphics[width=\textwidth,trim={3cm 0.6cm 9cm 0.4cm},clip]{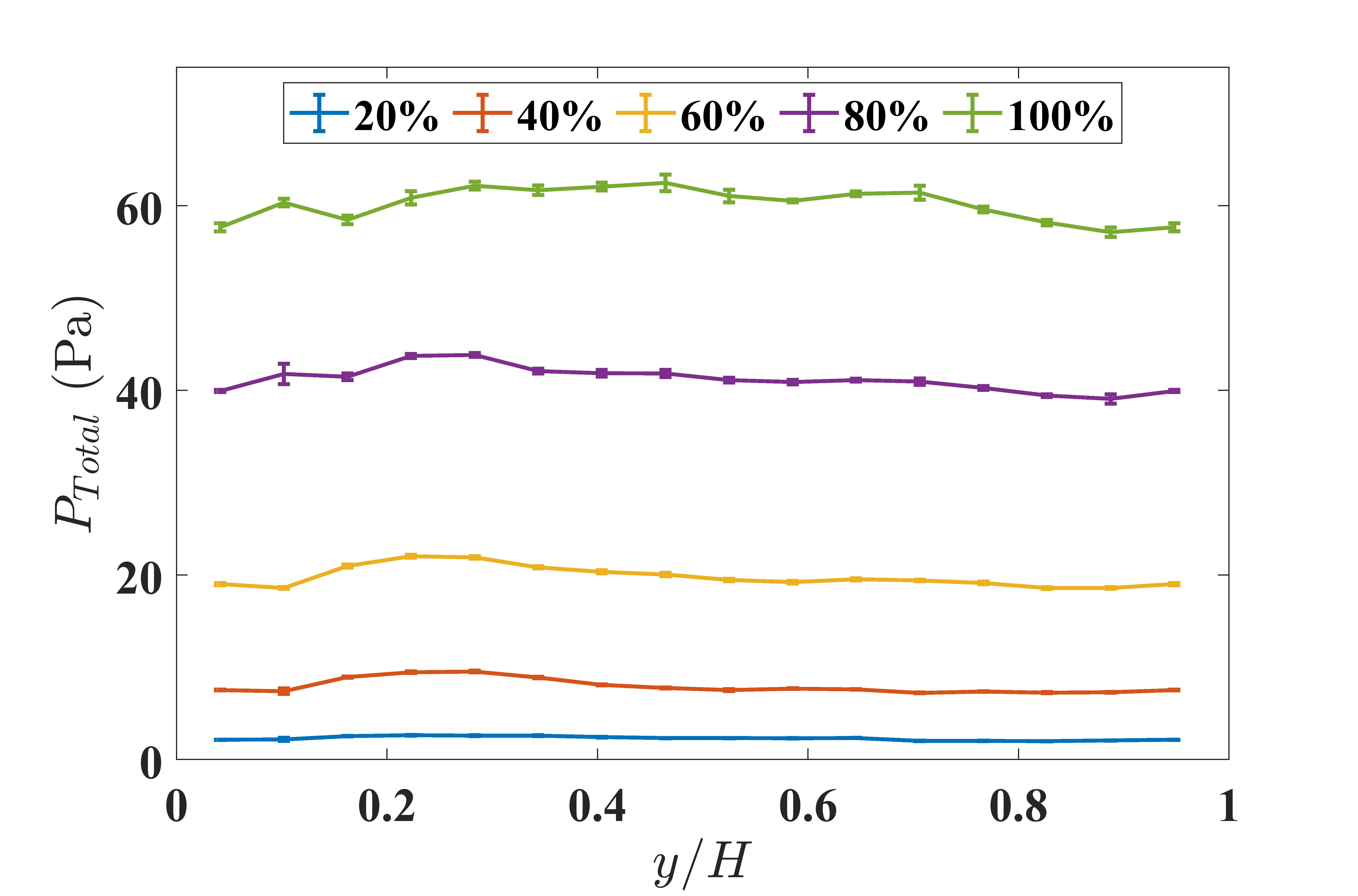}
\caption{ }
\label{fig:Totalpressure}
\end{subfigure}
\hfill
\begin{subfigure}[b]{0.49\textwidth}
\centering
\includegraphics[width=\textwidth,trim={2cm 0.6cm 9cm 0.4cm},clip]{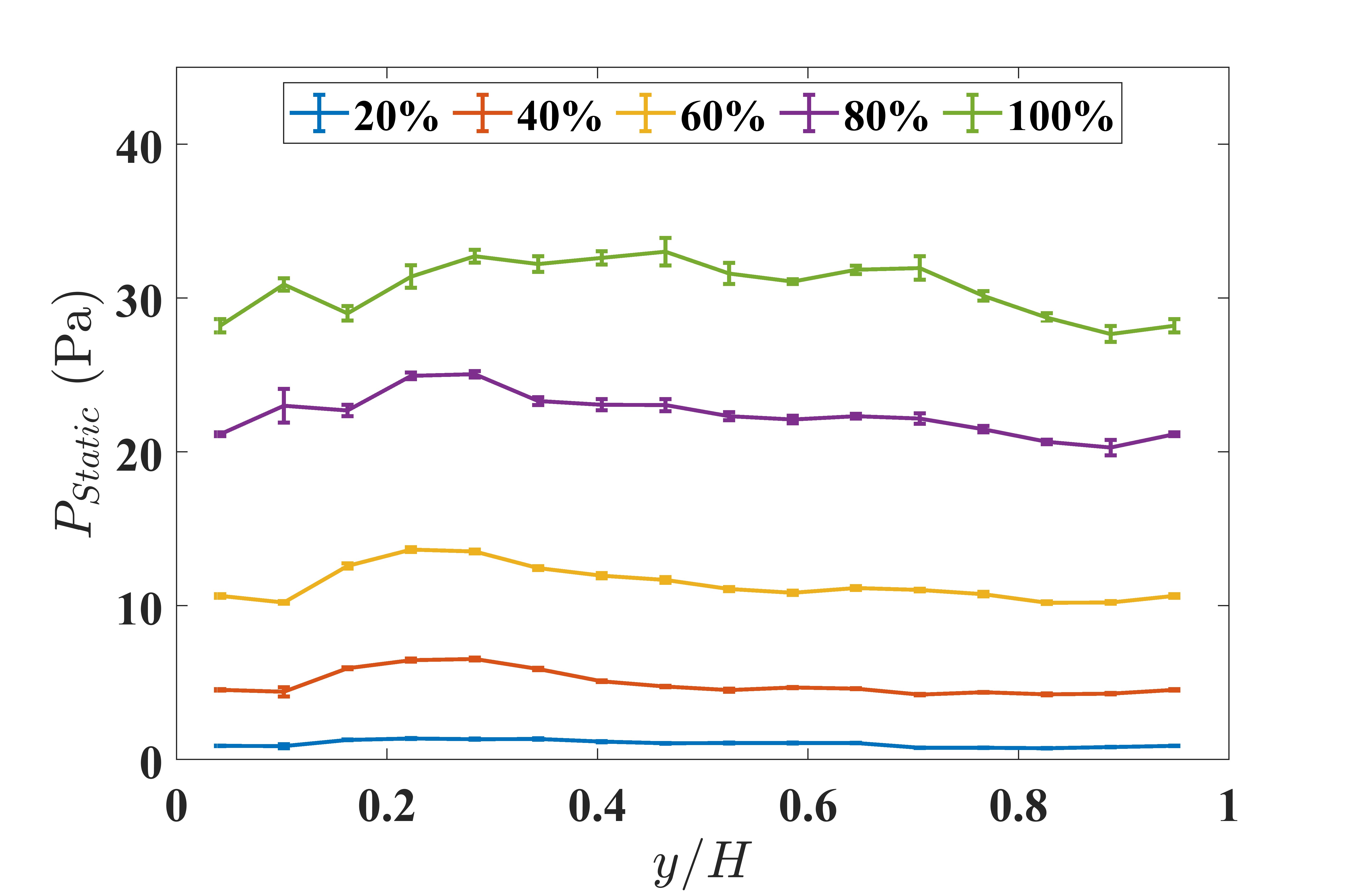}
\caption{ }
\label{fig:StaticPressure}
\end{subfigure}
\caption{Variation of (a) $P_{Total}$, and (b) $P_{Static}$ along $y/H$ for $x/L=0.52$ and $z/W = 0.5$.}
\end{figure}

\begin{figure}
\centering
\begin{subfigure}[b]{0.49\textwidth}
\centering
\includegraphics[width=\textwidth,trim={3cm 0.6cm 9cm 0.4cm},clip]{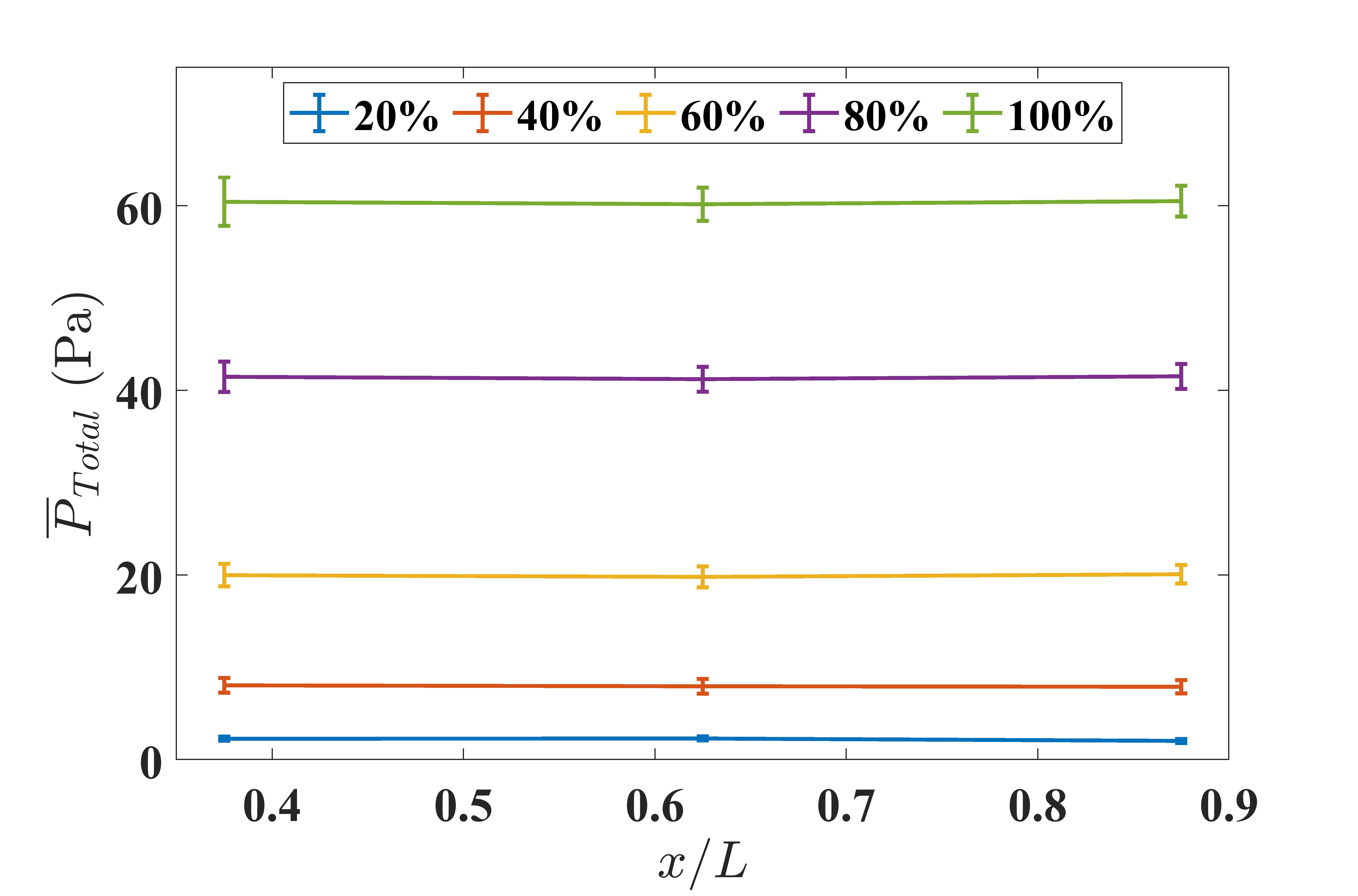}
\caption{ }
\label{fig:MeanTotalpressure}
\end{subfigure}
\hfill
\begin{subfigure}[b]{0.49\textwidth}
\centering
\includegraphics[width=\textwidth,trim={2cm 0.6cm 9cm 0.4cm},clip]{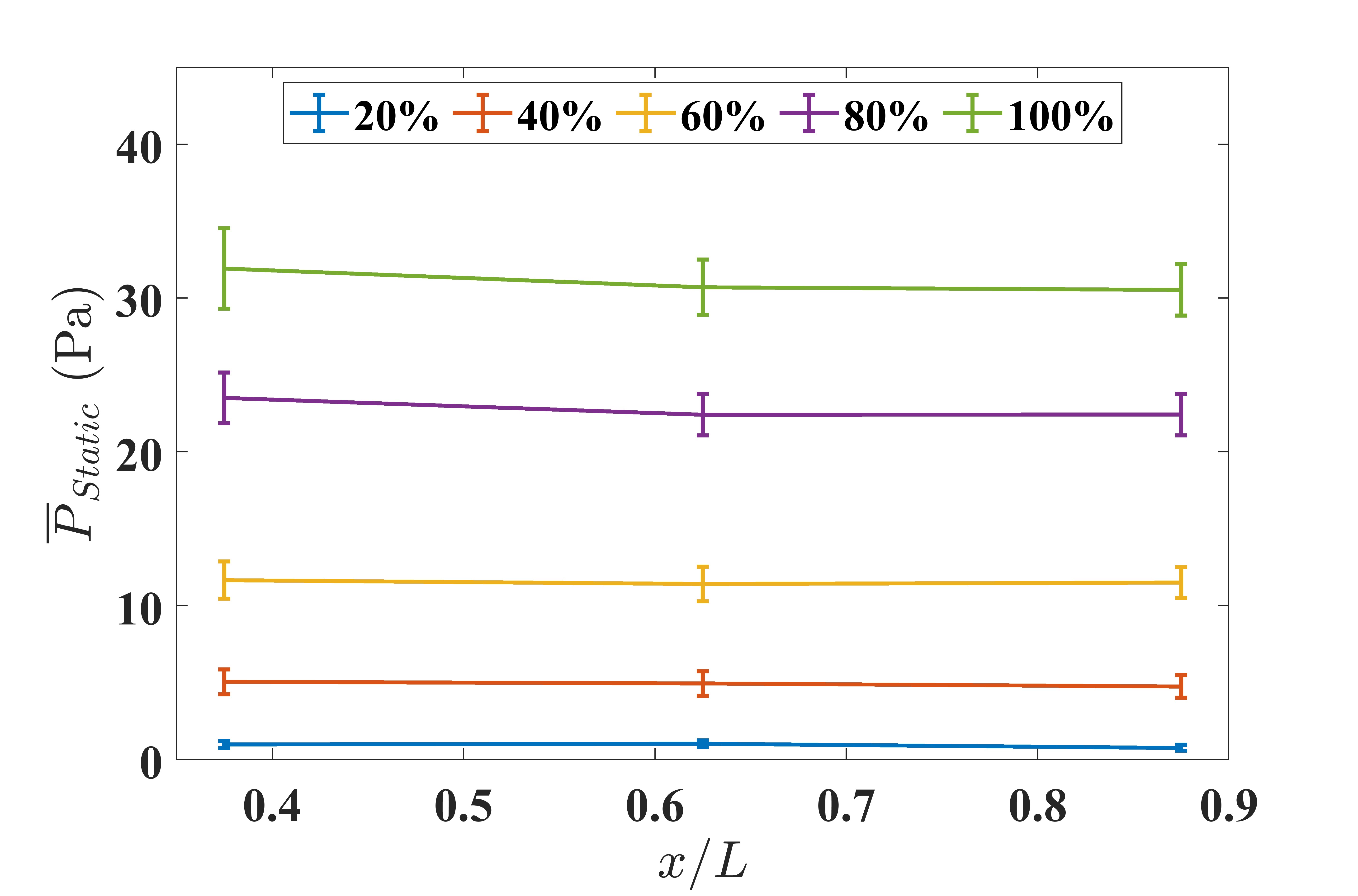}
\caption{ }
\label{fig:MeanStaticPressure}
\end{subfigure}
\caption{Variation of (a) $\bar{P}_{Total}$, and (b) $\bar{P}_{Static}$ along $y/H$ for $x/L=0.52$ and $z/W = 0.5$.}
\end{figure}

To understand the flow characteristics inside the tunnel, a 16-port pressure rake is considered to investigate the variation of total and static pressures along the FAWT’s test-section. Figures \ref{fig:Totalpressure} and \ref{fig:StaticPressure} present the variation of total ($P_{Total}$) and static ($P_{Static}$) pressures along the height at $z/W$ = 0.5 for a cross-section located at $x/L$ = 0.52 for duty-cycles $\in$ {20\%, 40\%, 60\%, 80\%, 100\%}. The pressure remains mostly the same along the height for all the duty-cycles investigated, and the variation in the $P_{Total}$ and $P_{Static}$ along the height is approximately 5 Pa for a 100\% duty cycle. To further understand the pressure distribution in the FAWT, Figures \ref{fig:MeanTotalpressure} and \ref{fig:MeanStaticPressure} present the variation of mean ($P_{Total}$)  ($\bar{P}_{Total}$)and mean $P_{Static}$ ($\bar{P}_{Static}$)along the length of the test-section. Here, $\bar{P}_{Total}$  and $\bar{P}_{Static}$ are defined as

\begin{equation}
\sum_{i=1}^n \frac{P_{\text {Total }, i}}{n} \text { and } \sum_{i=1}^n \frac{P_{\text {Static }, i}}{n}
\end{equation}

respectively, where n represents the number of ports along the height. From the figures, it can be observed that $P_{Total}$ and $P_{Static}$ remain uniform along $x/L$. Maximum pressure drop along the test-section is about 1.389 Pa for the 100\% duty-cycles.

\section{Linearly Sheared and Parabolic Profiles}

After investigating the flow characteristics for the uniform operating mode, the current section will present the flow characteristics for two non-uniform velocity profiles: (i) linearly sheared and (ii) parabolic velocity profiles.  

\subsection{Linearly Sheared flow profiles}

To generate a linearly sheared profile, the duty cycles of each row are varied. With the duty cycle of the bottom row being fixed at a 40\% duty cycle, all the rows starting from the bottom are operated at a 2\% and 4\% increment ($\Delta DC$) in duty cycles. For $\Delta DC = 2\%$, the duty cycles range from  $\in$ [40 - 58] \% starting from the bottom-most to the top-most row, respectively. While for $\Delta DC = 4\%$, the duty cycles range from  $\in$ [40 - 76] \%. To ensure repeatability, the data from the Hot-Wire Anemometer (HWA) were acquired over five repeated experimental runs, with a consistent sampling rate of 10 kHz.

\begin{figure}
\centering
\begin{subfigure}[b]{0.49\textwidth}
\centering
\includegraphics[width=\textwidth,trim={3cm 0.6cm 9cm 0.4cm},clip]{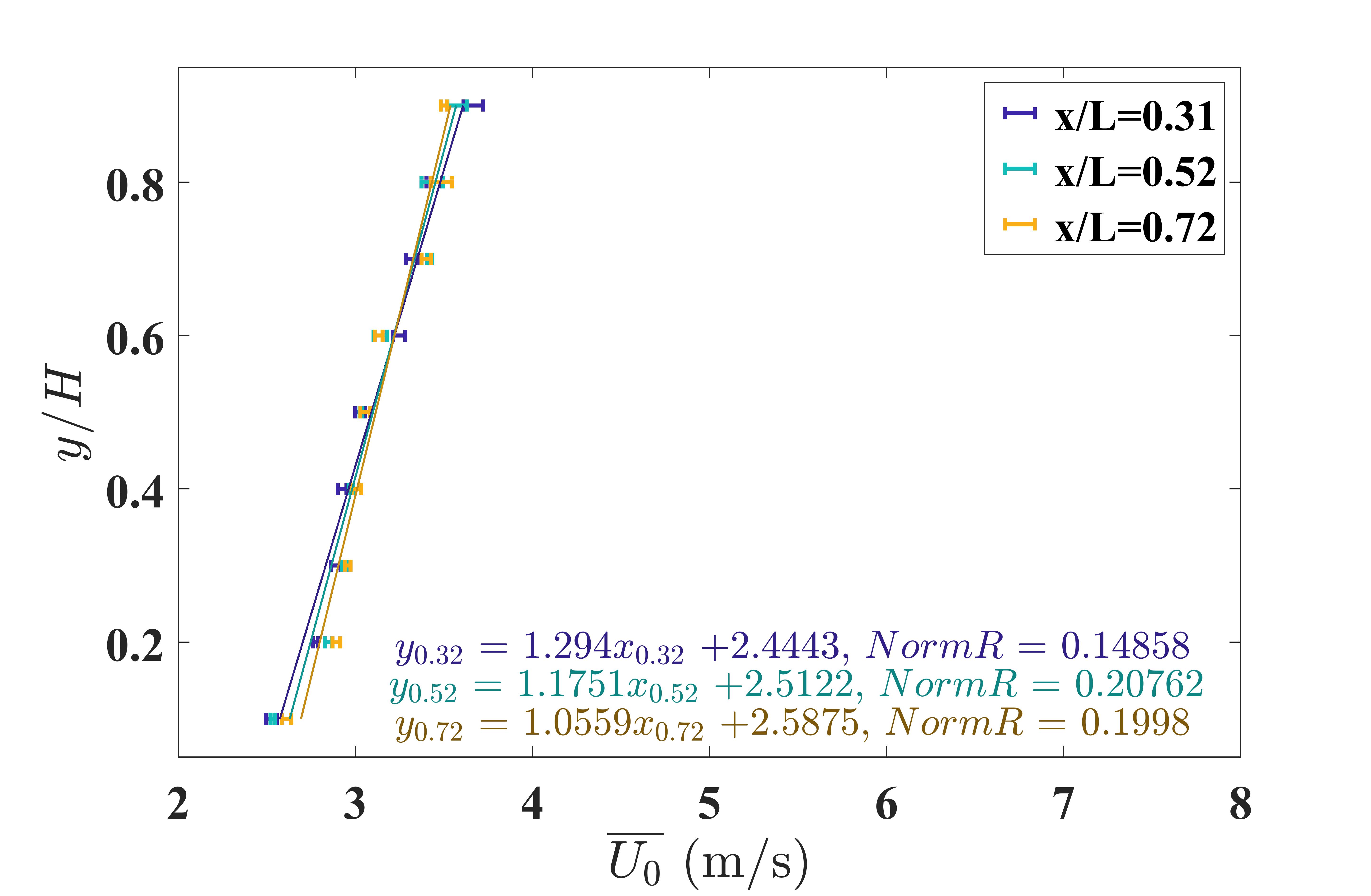}
\caption{ }
\label{fig:2pincreaselinear}
\end{subfigure}
\hfill
\begin{subfigure}[b]{0.49\textwidth}
\centering
\includegraphics[width=\textwidth,trim={2cm 0.6cm 9cm 0.4cm},clip]{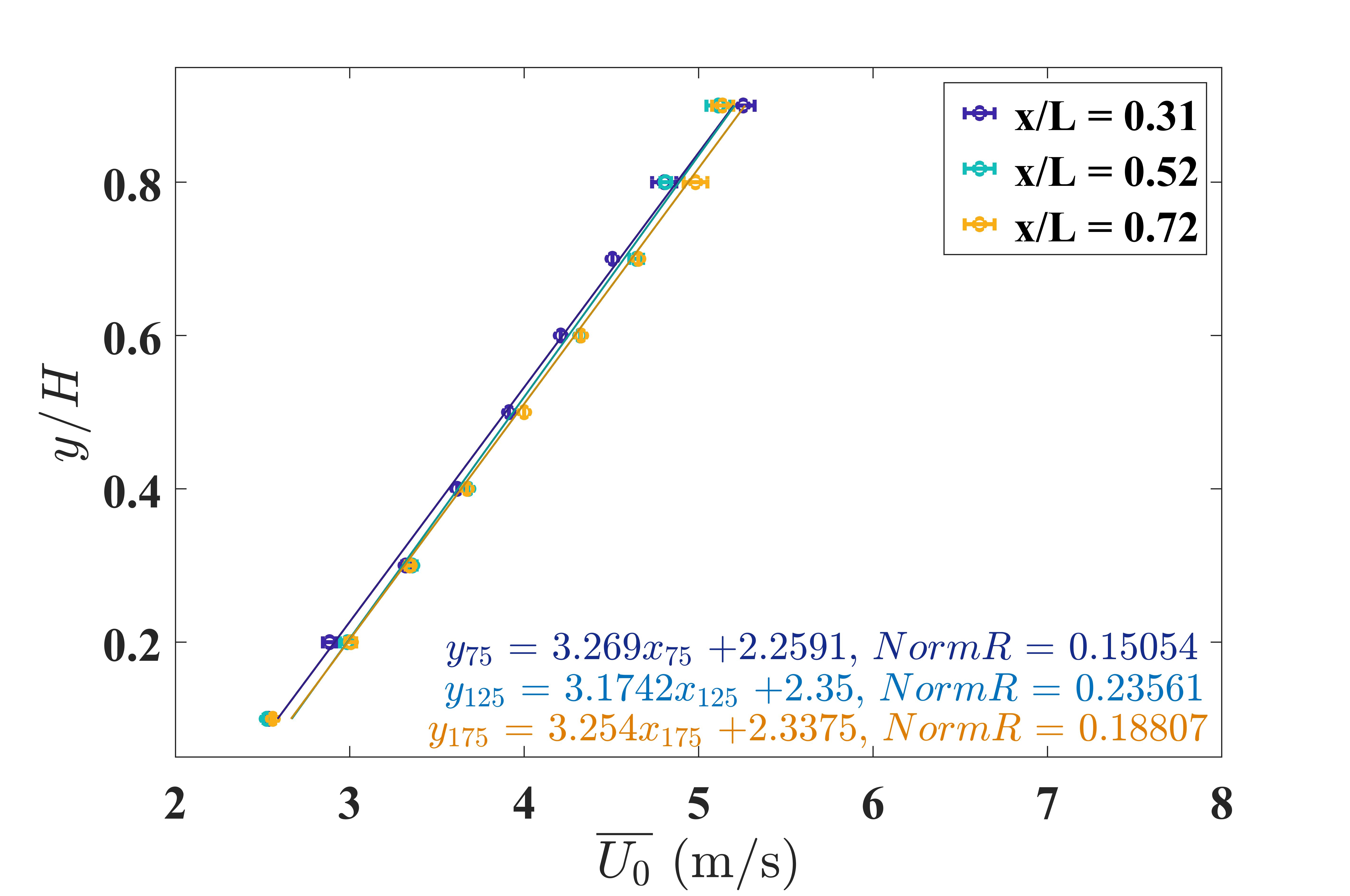}
\caption{ }
\label{fig:4pincreaselinear}
\end{subfigure}
\caption{Variation of $\overline{U_0}$ along $y/H$ at $x/L \in [0.31-0.72]$ for a linear increment of (a) $\Delta DC=$2\% (b) $\Delta DC=$4\%.}
\label{Umeanlinear}
\end{figure}

Figure~\ref{Umeanlinear} presents the variation of the mean shear flow velocity ($\overline{U_0}$ along the length of the test section for $x/L \in \{0.31,0.52,0.72\}$ for both cases. The figure clearly shows that the FAWT produces very good linearly sheared flows over a very short fetch length without any surface blocks with the aid of a microcontroller. A best-fit line is plotted through the data points to establish the trend followed by $\overline{U_0}$ along the non-dimensional height of the test section ($y/H$) while positioning the HWA at three different locations along the length of the cross-section as shown in figure\ref{Umeanlinear}. It can be observed that there is a very minimal change in the velocity profile along the length of the test section, and a higher $\overline{U_0}$ is observed for $\Delta DC = 4\%$. 

\begin{figure}
\centering
\begin{subfigure}[b]{0.49\textwidth}
\centering
\includegraphics[width=\textwidth,trim={3cm 0.6cm 9cm 0.4cm},clip]{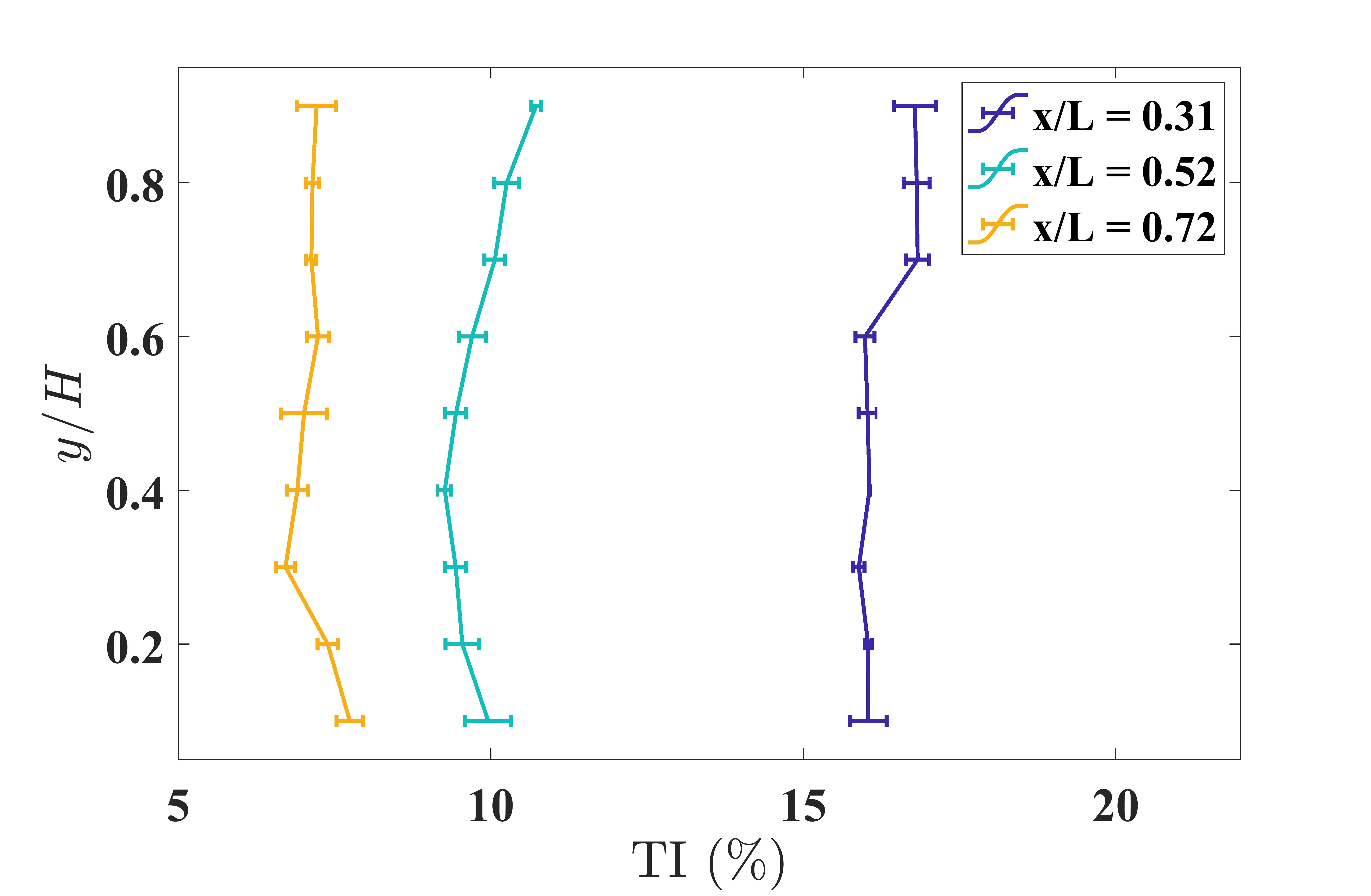}
\caption{ }
\label{fig:2pincreaselinear_TI}
\end{subfigure}
\hfill
\begin{subfigure}[b]{0.49\textwidth}
\centering
\includegraphics[width=\textwidth,trim={2cm 0.6cm 9cm 0.4cm},clip]{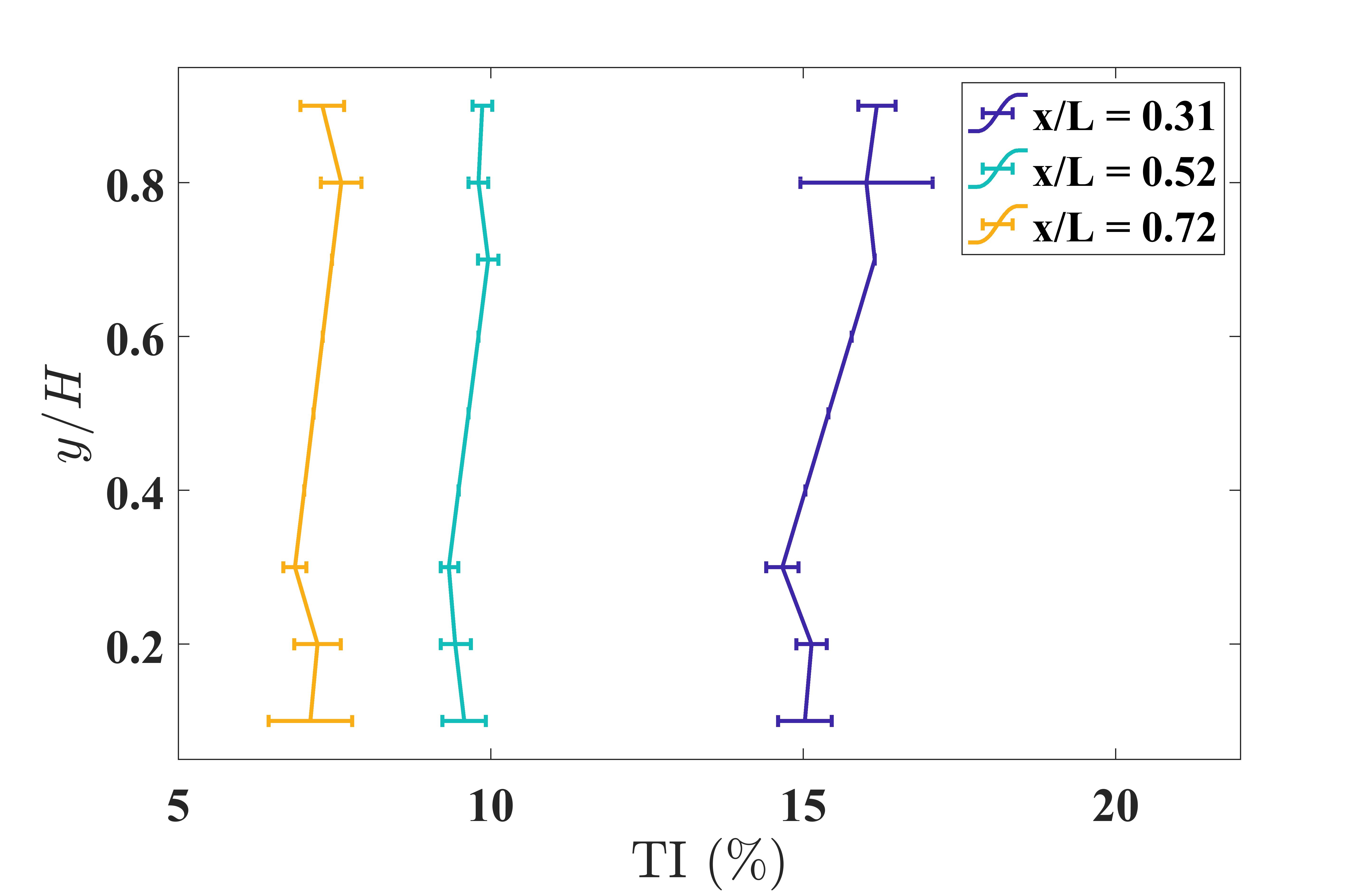}
\caption{ }
\label{fig:4pincreaselinear_TI}
\end{subfigure}
\caption{Variation of TI along $y/H$ at $x/L \in [0.31-0.72]$ for a linear increment of (a) $\Delta DC=$2\% (b) $\Delta DC=$4\%.}
\label{TILinear}
\end{figure}

Figure~\ref{TILinear} represents the turbulent fluctuations along the non-dimensional height of the test section ($y/H$) at various distances ($x/L$) from the fan-array along the length of the test section. It can be seen from the figures that the turbulent fluctuations remain uniform along the height of the test section. The TI is higher at locations nearer to the fan array, as already observed in the figure \ref{fig:TIvsx}, which can be attributed to the proximity of the sensor to the fan array configuration. As it moves further away, the flow stabilizes and the TI decreases. The trend remains consistent for both the duty cycle increments of $\Delta DC = 2\%$ and 4\%. 

\subsection{Parabolic flow profiles}

\begin{figure}
\centering
\begin{subfigure}[b]{0.49\textwidth}
\centering
\includegraphics[width=\textwidth,trim={3cm 0.6cm 9cm 0.4cm},clip]{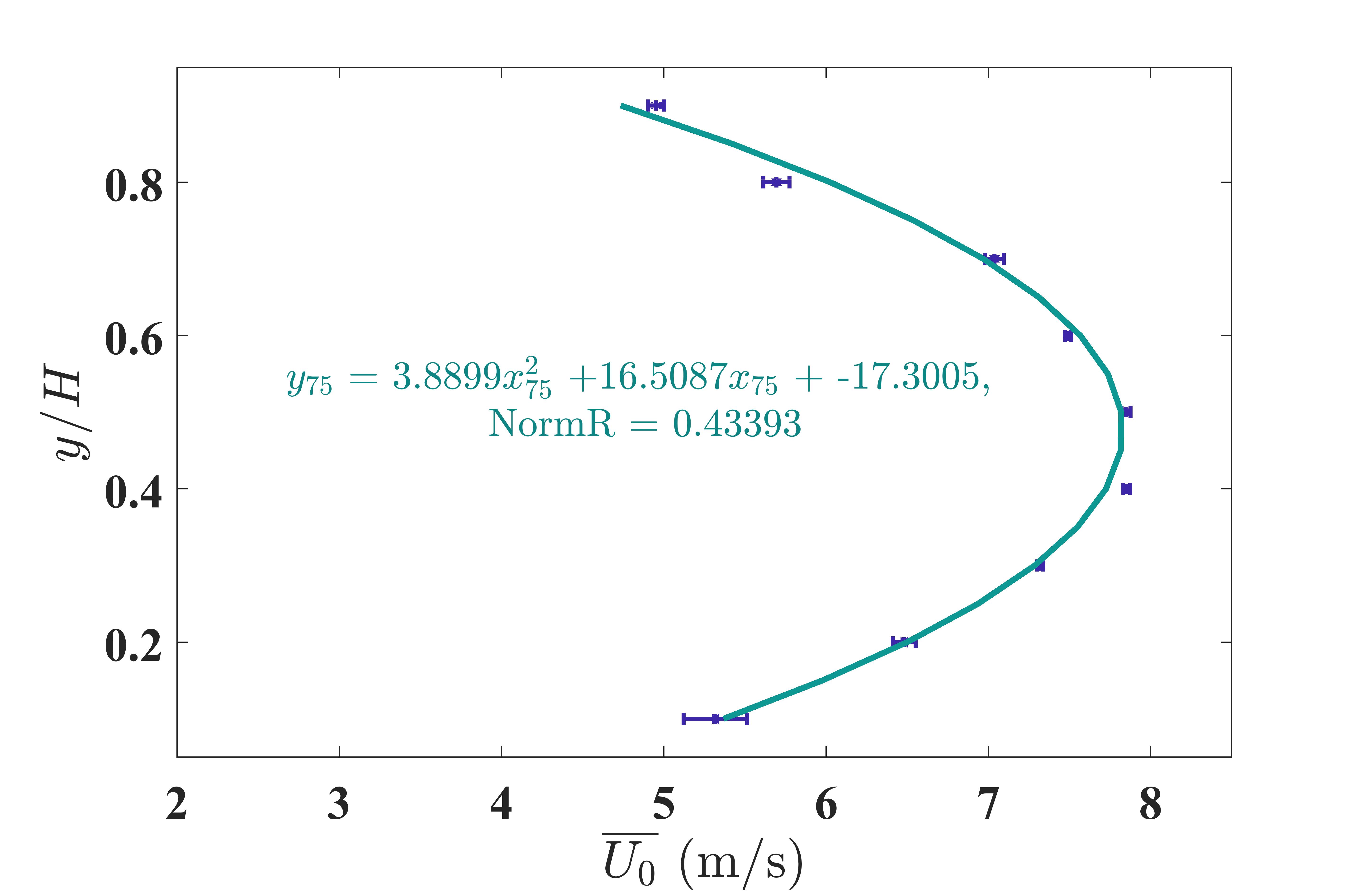}
\caption{ }
\label{fig:umeanparabolic}
\end{subfigure}
\hfill
\begin{subfigure}[b]{0.49\textwidth}
\centering
\includegraphics[width=\textwidth,trim={2cm 0.6cm 9cm 0.4cm},clip]{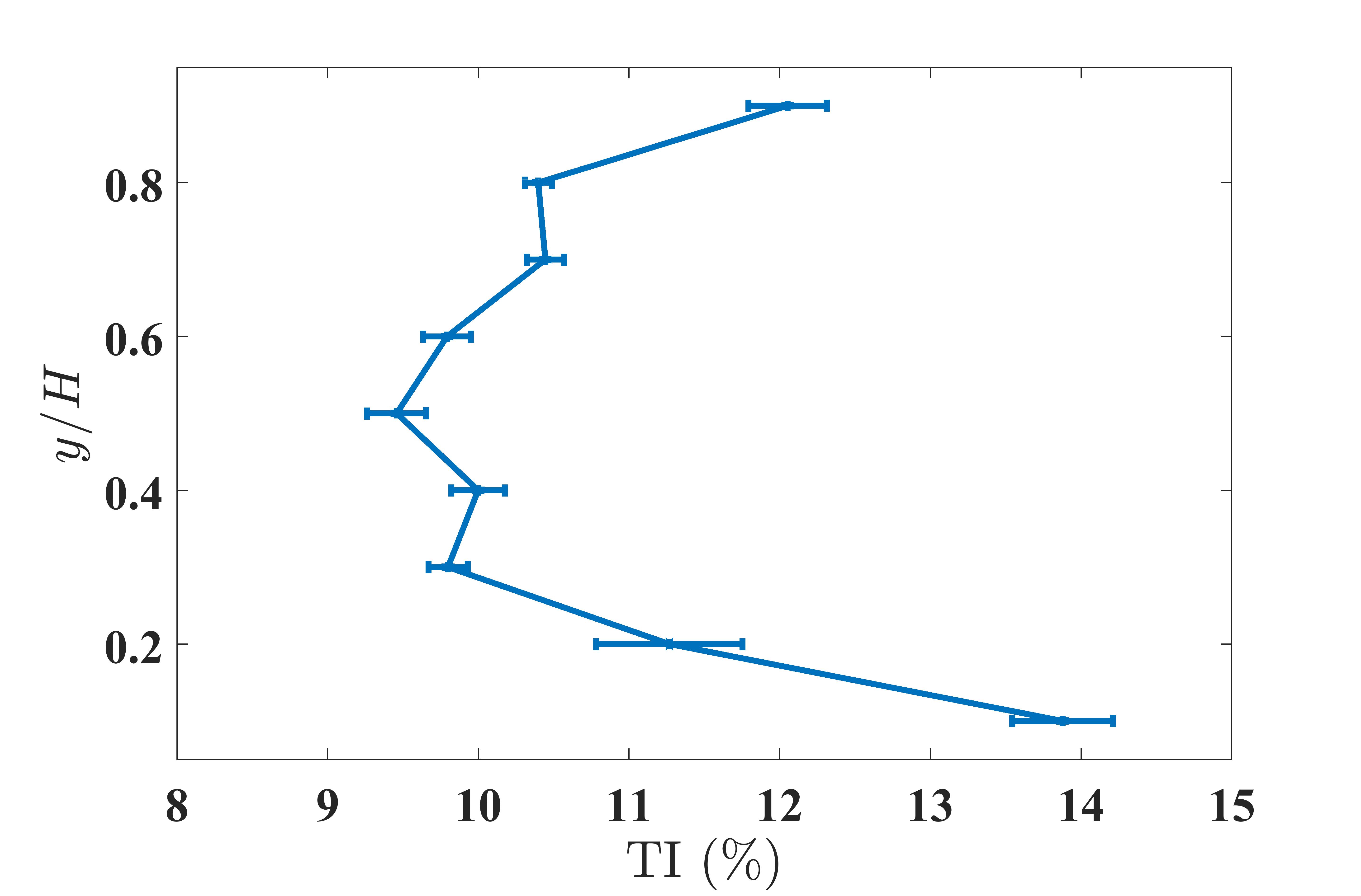}
\caption{ }
\label{fig:TIparabolic}
\end{subfigure}
\caption{Variation of (a) $\overline{U_0}$, and (b) TI along $y/H$ for $x/L=0.52$ and $z/W = 0.5$ when subjected to parabolic flow conditions.}
\end{figure}

To achieve a parabolic flow velocity profile, the duty cycle for each fan row was precisely determined by fitting a second-order polynomial equation (parabolic equation). The duty cycles range from $\in [0-96] \%$ along the 10 rows of the fan array, which is gradually increasing from 0\% to 96\% from the bottom row to the fifth row, and then gradually decreases from 96\% to 0\% from the sixth row to the topmost tenth row. Figure \ref{fig:umeanparabolic} shows the velocity distribution along the height of the test section($y/H$). The data is acquired at the center of the test section at a distance of 1.2 m from the fan array for a non-dimensional length of $x/L = 0.52$. It can be observed that the mean shear velocity ($\overline{U_0}$) follows a parabolic trend. $\overline{U_0} \in [5-8]$ m/s across the height of the test section ($y/H)$, with $\overline{U_0}$ being the maximum at the center of the test section.

Figure \ref{fig:TIparabolic} shows the TI variation along $y/H$ for the same case of the parabolic setting. It can be seen from the figure that the TI follows an inverse trend as compared to $\overline{U_0}$. As TI is the ratio of the root-mean-square (RMS) of the turbulent velocity fluctuations to the mean velocity, the TI is inversely proportional to the mean velocity ($\overline{U_0}$). Therefore at higher $\overline{U_0}$, the TI is lower. For the parabolic case, the TI ranges from $\in [9-14] \%$ as can be seen from the fig \ref{fig:TIparabolic}.

\section{Conclusions}
The work characterizes the performance evaluation of an FAWT subjected to uniform and non-uniform flow patterns. A 2.4 m long and 1.2 m wide fan-array wind tunnel was constructed, consisting of a 10 × 10 array of BLDC fans with a confined test section, and no flow modulating devices. At various duty-cycles ranging from 10\% to 100\% and at selected locations along the non-dimensional length ($x/L$), height ($y/H$) and width ($z/W$), flow parameters such as mean streamwise velocity ($\overline{U_0}$), turbulence intensity (TI), integral length scale ($l_s$), integral time scale ($\tau$), power spectral density (PSD), total pressure ($P_{Total}$), and static pressure ($P_{Total}$) were measured and evaluated. For turbulence measurements, a hot-wire anemometer was employed, and a multi-port pressure scanner was utilized to obtain the pressure data. The observations and findings are summarized below:-

1. The $\overline{U_0}$ increases linearly for duty- cycles $\in$ 30\% to 90\% with an approximate slope of 0.95 m/s per 10\% increase in duty-cycle. For a given point on the cross-section, $\overline{U_0}$ remained the same along the test-section. The $\overline{U_0}$ remains constant at a cross-section $x/L$ = 0.52 for $y/H= z/W =\{0.2 $-$ 0.8\}$. The boundary-layer effect is observed for $y/H$, $z/W$ $\le$ 0.2 and $y/H$, $z/W$ $\ge$ 0.8.

2. Turbulent intensity (TI) was observed to range between 2\% to 11\% across the test-section as a function of duty-cycle. It was observed that TI values remained similar for a cross-section at duty-cycles 50\% and above; however, for duty-cycles below 50\%, a substantial reduction in the TI values was noted for the same section. Additionally, the TI was observed to decrease along the length of the test-section due to increased turbulence kinetic energy transferring from the free-stream to the boundary layer with increasing boundary layer thickness.

3. The integral time scale ($\tau$) of the eddies was observed to be shorter with increasing duty cycle, and this phenomenon can be attributed to the increasing TI with duty cycle. Interestingly, ($\tau$) was observed to increase along the length test-section while the TI was reduced due to the energy injected into the eddies from the free-stream shear due to the fan-array. The integral length scale (ls) of eddies was observed to remain $\approx$ 2.5m/s at a cross-section for a range of duty-cycles. On the other hand, ls increased along the length of the test-section due to an increase in $\bar{\tau}$. The power spectral density (PSD) analysis of the turbulent kinetic energy exhibits the -5/3 slope for the inertial regime throughout the FAWT test-section for a range of duty-cycles.

4. The static pressure remained approximately uniform across the cross-section and along the length of the test-section for a duty cycle.

5. For the non-uniform operating modes, under linearly sheared flow conditions, the mean velocity ($\overline{U_0}$) and TI increases along the non-dimensional height ($y/H$), and the $\overline{U_0}$ remains similar along the length of the test section, but the turbulent fluctuations are higher at the location nearer to the fan-array at $x/L = 0.31$.

6. A parabolic profile for the $\overline{U_0}$ distribution is observed when the fan-array is operated on duty-cycles pre-determined by using a parabolic equation. An inverse parabola for the TI can be observed for this operating mode. 
    
The aerodynamic characteristics of an FAWT, under various flow conditions, have been extensively investigated in this work. Future extension of the current work can be the addition of turbulence grids and mesh blocks, to tune the turbulence intensities, to achieve desired wind conditions varying spatially and temporally for advanced aerodynamic research, as well as the development of machine learning based wind and gust generators.

\section*{acknowledgments}
The corresponding author would like to acknowledge the financial support from the Science and Engineering Research Board's Start-up Research Grant (SERB-SRG) with sanction order number SRG/2019/001249 and the BITS Pilani's OPERA award.

 \bibliographystyle{elsarticle-harv} 
 \bibliography{cas-refs}
\end{document}